\documentclass[]{aastex631}

\shorttitle{MHD simulations of flares}
\shortauthors{Rempel et al.}

\newcommand{\new}[1]{{#1}}

\renewcommand{\vec}[1]{\mathbf{#1}}

\graphicspath{{./}{figures/}}

\begin{document}

\title{Comprehensive Radiative MHD Simulations of \new{Eruptive} Flares above Collisional Polarity Inversion Lines}

\author[0000-0001-5850-3119]{Matthias Rempel}
\affiliation{High Altitude Observatory, National Center for Atmospheric Research, P.O. Box 3000, Boulder, CO 80307, USA}
\email{rempel@ucar.edu}

\author[0000-0002-1253-8882]{Georgios Chintzoglou}
\affiliation{Lockheed Martin Solar and Astrophysics Laboratory, 3251 Hanover Street, Palo Alto, CA 94304, USA}

\author[0000-0003-2110-9753]{Mark C. M. Cheung}
\affiliation{CSIRO, Space \& Astronomy, PO Box 76, Epping, NSW 1710, Australia}

\author[0000-0003-1027-0795]{Yuhong Fan}
\affiliation{High Altitude Observatory, National Center for Atmospheric Research, P.O. Box 3000, Boulder, CO 80307, USA}

\author[0000-0002-7791-3241]{Lucia Kleint}
\affiliation{Astronomical Institute of the University of Bern, Sidlerstrasse 5, 3012 Bern, Switzerland}

\begin{abstract}
We present a new simulation setup using the MURaM radiative Magnetohydrodynamic (MHD) code that allows to study the formation of collisional polarity inversion lines (cPILs) in the photosphere and the coronal response including flares. In \new{this scheme} we start with a bipolar sunspot configuration and set the spots on collision course by imposing the appropriate velocity field at the footpoints in the subphotospheric boundary. We \new{produce different setups with the same initial spot separation by varying physical} parameters such as the collision speed and \new{minimum} collision distance. While all setups lead to the formation of an \new{EUV and X-ray} sigmoid structure, only the cases with a close passing of the spots cause flares and mass eruptions. The energy release is in the $1-2 \times 10^{31}$ ergs range, putting the simulated flares into the \new{upper C-class to lower M-class range of GOES X-ray 1-8\,\AA\ flux}. While the \new{setup} with the more distant passing of the spots does not lead to a flare, the corona is nonetheless substantially heated, suggesting non-eruptive energy release mechanisms. We focus our discussion on two \new{particular} setups that differ in spot coherence and resulting cPIL length \new{persistence}. We find different timings in the transition from a sheared magnetic arcade (SMA) to magnetic flux rope (MFR); the setup with a \new{large length but} shorter \new{duration} cPIL produces a MFR during the eruption, while the MFR is pre-existing in the setup with a \new{large length} \new{and} longer \new{duration} cPIL. While both result in flares of comparable strength \new{and the eruption of a CME, the setup with pre-existing MFR (and embedded filament) leads to an MFR eruption with a larger mass content.}
\end{abstract}


\section{Introduction} \label{sec:intro}

Based on observations there is solid statistical evidence that solar Active Regions (ARs) with complex polarity inversion lines (PILs; the imaginary line separating opposite-signed polarities) are very flare- and CME-productive \citep{Schrijver:2007}. Statistical flare forecasting approaches rely on various measures of AR complexity\new{, such as PIL length, field gradients across PIL, vertical electric current strength, shear angle of the observed  photospheric horizontal magnetic field (measured with respect to the potential horizontal field along the PIL)} and generally find a higher likelihood for flares in more complex ARs \new{(e.g., \citealt{Gallagher:etal:2002,Yanmei:etal:2006,Yanmei:etal:2007,Barnes:etal:2007,Georgoulis:Rust:2007})}. The most flare-productive ARs are often associated with $\delta$-spots \citep[classification introduced by][]{Kuenzel:1960}, which are characterized by sunspot umbrae of opposite magnetic polarities sharing a penumbra \new{ -- a manifestation of opposite magnetic polarities in very close proximity to each other}.

A significant effort in Magnetohydrodynamic (MHD) modeling has focused on understanding how $\delta$-spots form and how they energize the corona and led to strong flares. While \cite{Linton:etal:1998,Linton:etal:1999,Takasao:etal:2015} studied the formation of $\delta$-spots through the emergence of twisted, kink-unstable flux tubes, \citet{Toriumi:etal:2014,Fang:Fan:2015} showed that flows during flux emergence, resulting from mass drainage, can pin down sections of a rising flux-bundle and lead to more complex magnetic field distributions in the photosphere. A more systematic study by \citet{Toriumi:Takasao:2017} (following \citealt{Zirin:Liggett:1987}) investigated four possible scenarios\new{, namely: (a) ``spot-spot'' interaction from a kink-unstable flux tube, (b) ``spot-satellite'' interaction, (c) quadrupolar setup from convective (or buoyant) down-pinning of a rising flux bundle, and (d) quadrupolar setup from the interaction of two rising flux bundles. The latter two scenarios were also used in \citet{Toriumi:etal:2014} to simulate the formation of NOAA AR11158, a very flare-productive AR \new{that produced extreme flaring and CME activity in February 2011}. While the setup they picked as the most plausible model (scenario (c) mentioned above) for the formation of such ARs produced large magnetic shear angle at the PIL at the photosphere, the motions of the colliding polarities forming that PIL did not reproduce the phenomenology of AR11158 as described in \citet{Chintzoglou:Zhang:2013, Chintzoglou:etal:2019}. In the observations, the colliding opposite-signed polarities slide past each other and continued on a trajectory to meet their like-signed polarities (e.g., see \citealt{Chintzoglou:Zhang:2013} for a 3D subphotospheric model) thereby increasing the magnetic shear, while in the \citet{Toriumi:etal:2014} and \citet{Toriumi:Takasao:2017} simulation, the proper motions of the opposite-signed polarities completely stopped during the closest encounter of the collision. Essentially, this complete stop of proper motions happened because, in their model, the two flux bundles were part of the same subphotospheric flux-tube, and the final position of the colliding polarities was prescribed due to the fixed, ``dipped'' setup in the sub-photosphere, which effectively pins down the two colliding polarities above the location of the dip, preventing them to move any more forward. \citet{Chintzoglou:etal:2019} pointed out that the observations suggest that the magnetic flux tubes which formed AR11158 (and other similar ARs) were not connected via a dip, but rather they behaved as separate, or, \emph{independent} magnetic bipoles, as evidenced by the polarity proper motions that persisted for a large part of the evolution during, and after, the emergence phase of the AR.}

\citet{Chintzoglou:etal:2019} unified the multitude of scenarios presented in \citet{Zirin:Liggett:1987} and \citet{Toriumi:Takasao:2017} by identifying a general process in which AR-forming flux bundles (typically two) emerge in close proximity (either simultaneously or after each other), leading to a quadrupolar magnetic field configuration. In such quadrupolar configuration, the opposite-signed non-conjugated polarities (i.e., polarities belonging to different bipolar groups within the AR) interact with each other through a process called \emph{``collisional shearing''} (i.e., polarities collide and shear and move past each other as at least one of the interacting bipoles emerges and drives the photospheric evolution and explosive activity in the corona). \new{Thus, various degrees of complexity in terms of time-evolving magnetic configurations can be achieved in solar ARs with the occurrence of subsequent events of emergence of bipoles during the initial formation phase of ARs (or even during the AR decay phase; see examples and discussion in \citealt{Chintzoglou:etal:2019}).}

While the aforementioned MHD simulations of \citet{Toriumi:Takasao:2017} focused on the formation of $\delta$-spots in the photosphere, their work did not study how the $\delta$-spot formation influences the buildup and release of free energy in the overlying corona leading eventually to energetic flares. This was studied by \citet{Chintzoglou:etal:2019} with the help of 3D time-evolving magneto-frictional modeling, showing how the the delta-spot led to the formation of a pre-eruptive \new{Magnetic Flux Rope (MFR). A MFR is a bundle of twisted magnetic field lines manifesting the accumulation of magnetic free energy in the corona above the region of collision of polarities. The same work by \citet{Chintzoglou:etal:2019} also showed that recurrent flaring and CME activity was produced preferentially above PIL locations undergoing collisional shearing.}

One critical aspect of the coronal magnetic field evolution concerns the question of how and when a MFR forms during the energy buildup and eruption process. It was proposed by \citet{Antiochos:etal:1999,Lynch:etal:2008} that MFRs form ``on-the-fly'' during the eruption process through resistive MHD processes at the reconnection site. The alternative scenario is a pre-existing MFR that formed during the energy buildup phase and destabilizes due to ideal MHD instabilities \citep[e.g.,][]{Kliem:Torok:2006}. These two scenarios are not mutually exclusive, since a pre-existing MFR can be enhanced during the eruption, and the pre-eruption configuration can exist in a hybrid state with a smooth transition from a sheared magnetic arcade (SMA) to a MFR \citep{Patsourakos:etal:2020}. \new{\citet{Chintzoglou:etal:2019} found that collisional shearing can form pre-eruptive structures as a result of the SMA-to-MFR conversion, the latter being due to systematic magnetic cancellation supported \emph{during the emergence phase} of complex ARs.}

The main focus of this paper is the study of the evolution of the coronal magnetic field during the process of collisional shearing from the early stages of energy buildup to eruption \new{via radiative convective 3D MHD simulations. Central to this process is the formation of a \emph{collisional Polarity Inversion Line} (cPIL). \citet{Chintzoglou:etal:2019} defined a cPIL as the AR's PIL segment that is formed by opposite-signed magnetic flux elements ($|B_z|\geq$100\,G) that are separated by a small maximum distance $\leq$ 1.8\,Mm (e.g., see similar studies by \citealt{Liu:etal:2019,Liu:etal:2021,Wang:etal:2022}). Most ARs undergoing collisional shearing include at least two interacting bipolar groups (i.e., a quadrupolar setup) where each bipole has its own ``self-PIL'', or in other words, its own \emph{internal} PIL (meaning the AR's PIL that separates \emph{conjugated} polarities, i.e., the positive and negative polarities belonging to the same emerging bipole). A cPIL forms in-between \emph{non-conjugated} polarities (i.e., opposite polarities from different bipolar groups), and thus it is an \emph{external} PIL from the perspective of each participating bipole. With regards to the quadrupolar AR as a whole, a cPIL could form in the AR's internal PIL; otherwise it could be part of an AR's external PIL, e.g., if collision happens with flux originating from another neighboring (emerging or decaying) AR. Here, we simplify our simulation setup to only consider the two (non-conjugated) interacting polarities, essentially by disregarding the conjugated polarities that are situated away from the collision region. In fact, rapidly-converging proper motions between the two large-scale polarities of a single bipole do not happen during the bipole's emergence phase; instead, diverging proper motions develop due to the characteristic ``self-separation'' of the two polarities about the internal PIL of the simple bipole (see illustration in  \citealt{Chintzoglou:etal:2019} Figure 1 and discussion therein).} We emphasize that the evolution presented here shall be understood by always keeping the two intentionally-omitted non-colliding polarities in mind, in accordance with the collisional shearing process. Furthermore, our 3D model setup is not to be confused with the \new{2.5D}\footnote{\new{2.5D implies a 3D setup in which the physical quantities are invariant along the dimension that is set to be aligned with the PIL.}} \new{magnetic flux cancellation scenario of \citet{Ballegooijen:Martens:1989} that considers a decaying  bipole, and the similarly bipolar 3D flux cancellation model of \citet{Amari:etal:2000} , because: (a) there is no consideration for proper motions for the bipole's polarities in such models, by construction; also, (b) cancellation in those bipolar models happens in the \emph{internal} PIL as a result of flux diffusion, and the shear is produced via differential rotation or even via arbitrary analytical flow profiles.} In our setup, the high speed of the moving polarities is inspired by the commonly observed rapid relative polarity motions that are typical during the emergence phase of any simple bipole (or between groups of bipoles); collision happens in the \emph{external} PIL of such separate bipoles, with the group of bipoles composing a single, complex multi-polar AR (see Figures 12 and 15 in \citealt{Chintzoglou:etal:2019}). Furthermore, we stress that our setup removes complexities originating from \new{active flux emergence (which take place in other MHD models of flux emergence, e.g., \citealt{Fan:2001, Cheung:etal:2019:HGCR})}, as we simply translate the magnetic polarities at their sub-photospheric roots, while still permitting cancellation of magnetic flux at the photosphere.

We describe the model setup in section \ref{sec:sim}, present the results in section \ref{sec:results} and discuss future observations that could directly differentiate between SMAs and MFRs through magnetic observations in section \ref{sec:future-obs}. We summarize our conclusions in section \ref{sec:concl}.

\section{Simulation Setup} \label{sec:sim}
For the simulations presented here we use the Coronal extension of the MURaM code \citep{Rempel:2017:corona} that has been successfully used for solar flare simulations as reported in \citet{Cheung:etal:2019:HGCR}. We use a domain \new{of $N_x\times N_y\times N_z = 512\times 256\times 1152$ (grid points)$^3$ with the extents of $L_x\times L_y\times L_z = 98.304\times 49.152\times 73.728$ Mm$^3$ ($\hat{z}$-direction points against gravity) with a grid cell size of $dx\times dy\times dz = 192\times 192\times 64$ km$^3$}. \new{The location of the photosphere is at about $z=h_{phot}\approx3.4$ Mm above the bottom boundary condition.} The domain is periodic in the horizontal directions. \new{Technically, such periodic boundary condition renders the setup not exactly bipolar but rather an infinite lattice in x-y space of bipolar setups. However, the interaction due to the collision happens locally about the cPIL, thus we do not see much influence from the periodic boundaries. This was further investigated for a comparable setup in \citet{Cheung:etal:2019:HGCR} by complementary magneto-frictional simulations using periodic and non-periodic  boundary conditions.} The top boundary is open for vertical flows and the magnetic field is matched to a potential field extrapolation. Outside regions of strong magnetic field the bottom boundary condition is open as described in \cite{Rempel:2014:SSD}. 
\new{The initial magnetic field configuration consists of a pair of spots with a separation of $24.576$ Mm, this initial state was created as follows: We started with a $4.096$~Mm deep photospheric simulation domain and inserted axisymmetric, self-similar magnetic field configurations \citep[see, Appendix A in][]{Rempel:2012:penumbra} at the grid positions (36.864~Mm, 24.576~Mm) and (61.44~Mm, 24.576~Mm) with opposite polarity. For the self-similar field we used a radial profile $f=\exp{[-(\zeta/R_0)^4]}$ with $R_0=2.88$~Mm and a vertical profile of $g=\exp{[-z/z_0]}$ with $z_0=1.257$~Mm, here $\zeta=R/\sqrt{g(z)}$ describes the self-similar expansion with height. With an initial field strength value of $B_0=13$~kG at the bottom boundary, this corresponds to a drop to $500$~G at the top boundary. We added this magnetic field to a statistically relaxed quiet Sun small-scale dynamo simulation and evolved the photospheric setup for 3 hours in order to allow the photospheric sunspot structure to settle in and form a Wilson depression. This resulted in sunspots with about $10-11$~kG field strength at their footpoint and $3-4$~kG in the photosphere and a magnetic flux of $3\times 10^{21}$ Mx for each spot. After this initial photospheric relaxation we added a corona following \citet{Rempel:2017:corona} by filling the coronal volume with a potential field and an isothermal stratification with $100,000$~K. This setup was evolved for another 3 hours until the corona reached a statistically relaxed state as described in \citet{Rempel:2017:corona}. We used this setup as the initial state for all the following simulations that differ only in terms of the bottom boundary condition as described below.} At the footpoints of the sunspots we impose within a circle $\sqrt{(x-x_c(t))^2+(y-y_c(t))^2}<R=4\,{\rm Mm}$ (we follow here the convention that the indices $-2$ and $-1$ correspond to the boundary cells, whereas the lowermost domain cells are $0$ and $1$):
\begin{eqnarray}
    \vec{M}_{-1}&=&\vec{M}_0 \,a_{M}+\varrho_{-1}\vec{v}_{\rm BND}\,(1-a_M)\nonumber\\
    \vec{M}_{-2}&=&\vec{M}_1 \,a_{M}+\varrho_{-2}\vec{v}_{\rm BND}\,(1-a_M)\label{Mbnd}\\
    \vec{B}_{-1}&=&\vec{B}_0 \,a_{B}\nonumber\\
    \vec{B}_{-2}&=&\vec{B}_0 \,a_{B}\label{Bbnd}\\
    p_{-1}&=&\bar{p}_0\,a_{p}+p_0^{\prime}\nonumber\\
    p_{-2}&=&\bar{p}_0\,a_{p}^2+p_1^{\prime}\label{pbnd}
\end{eqnarray}
Here $\vec{M}$ and $\vec{B}$ are the vectors of mass flux and magnetic field, $p$ and $\varrho$ are gas pressure and density, $\bar{p}$ denotes the horizontally averaged pressure and $p^{\prime}$ the pressure perturbation. The coefficients \new{$a_M$}, $a_B$, $a_p$ are given by
\begin{equation}
    a_M=0.8,\hspace{0.5cm}
    a_B=\vert\vec{B}_0\vert / {\rm max}(0.8\vert\vec{B}_0\vert,\vert\vec{B}_1\vert),\hspace{0.5cm}
    a_p=\bar{p}_0/\bar{p}_1\label{coeff}
\end{equation}
and the imposed boundary velocity is related to the motion of the footpoint $(x_c, y_c)$ through:
\begin{equation}
    \vec{v}_{\rm BND}=(\varepsilon_h\,\dot{x}_c,\varepsilon_h\,\dot{y}_c,0)
\end{equation}
\new{The circular region in which we impose the above boundary condition moves with the velocity $(\dot{x}_c,\dot{y}_c)$. The parameter $\varepsilon_h$ allows us to impose within that circular region a velocity that differs slightly from $(\dot{x}_c,\dot{y}_c)$, which impacts the coherence of the footpoint. A choice of $\varepsilon_h<1$ leads to steady flux loss (footpoint moving slower than the circular region), while $\varepsilon_h>1$ minimizes flux loss by pushing the flux towards the leading edge of the circular region $\sqrt{(x-x_c(t))^2+(y-y_c(t))^2}<R$ with respect to the flow direction (while the footpoint is pushed towards the leading edge, it cannot move faster than the circular region in which we impose this boundary condition). In the equation for the mass flux boundary Eq. (\ref{Mbnd}) $a_M$ is a coupling parameter which is used to minimize artifacts from imposing the velocity directly. While a choice of $a_M=0.8$ allows for short term deviations from $\vec{v}_{\rm BND}$, it enforces $\vec{v}_{\rm BND}$ over longer time-scales (more than 10-20 iterations).} The magnetic field Eq. (\ref{Bbnd}) is extrapolated into the boundary cells using the gradient of the magnetic field strength present in the lowermost domain cells, given by $a_B$. The formulation of $a_B$ from Eq. (\ref{coeff}) prevents extreme values that would lead to numerical instability. The pressure boundary condition corresponds to an extrapolation of the mean stratification and a symmetric boundary for pressure fluctuations.

For the simulations presented here we held the left sunspot in place, i.e. $\vec{v}_{\rm BND}=0$ and $(x_c,y_c)=(36.864\,{\rm Mm}, 24.576\,{\rm Mm})$, while we move the right sunspot according to the parameters as given in Table~\ref{tab:tab1} starting from the position $(x_c,y_c)=(61.44\,{\rm Mm}, 24.576\,{\rm Mm})$. For $t<t_{\rm acc}$ we increase the footpoint velocity linearly from $0$ to $(\dot{x}_c,\dot{y}_c)=(v_x^0,v_y^0)$, for $t\geq t_{\rm acc}$ the velocity remains constant. In all simulations we use $t_{\rm acc}=10^4$ s.

\begin{table}
\begin{center}
    \begin{tabular}{c|c|c|c|c|c|c|c|c|c|c}
    \hline\hline
        Setup & $v_x^0$ & $v_y^0$ & $\varepsilon_h$ & Collision speed & $t_{min}$ & $l_{min}$ & $L_{cPIL}^{min}$ & $t_{peak}$ & $l_{peak}$ & $L_{cPIL}^{peak}$ \\
         & [m/s] & [m/s] & [unitless] & [m/s]  & [hours] & [Mm] & [Mm] & [hours] & [Mm] & [Mm]\\ \hline
        A  & -954 & 300 & 1.2 & 1000 &  7.6 & 8.6 & 19.3 & 9.8 & 10.4 & 7.6 \\
        B\footnote{Setup B has been also used in \citet{Cheung:etal:2022}} 
  & -954 & 300 & 1   & 1000  & 9.5 & 8.5 & 20.0 & 13.8 & 11.2 & 11.8 \\
        C  & -917 & 400 & 1.2 & 1000  & 7.6 & 10.1 & 9.0 & 11.5 & 17.0 & 6.6 \\
        D  & -866 & 500 & 1.2 & 1000 &  7.5 & 12.1 & 1.7 & 14.2 & 28.6 & 0 \\
        E  & -458 & 200 & 1.2 & 500 & 13.9 & 10.0 & 2.9 & 21.7 & 17.0 & 0 \\ 
    \end{tabular}   
    \caption{\label{tab:tab1} Parameters for boundary driving used in the simulations \new{and measured time and value of collision distance and collisional PIL length at the time of minimum collision distance, $t_{min}$. Additional columns show the collision distance and collisional PIL length at the peak time of flare, $t_{peak}$ (or the peak time of the strongest flare, in case more than one flare was produced).} Runs (A)-(D) have a sunspot moving with $1000$ m/s, run (E) with $500$ m/s. Run (B) uses a parameter of $\varepsilon_h=1$, which leads to a dispersal of the moving sunspots. In all other runs using $\varepsilon_h=1.2$ the moving sunspot remains coherent throughout the duration of the simulations. The sequence of run (A), (B), (C), and (D) leads to increasing collision distance, while runs (C) and (E) have similar collision distance but different collision speeds. }
\end{center}
\end{table}

\section{Results} \label{sec:results}
\begin{figure}
    \centering
    \resizebox{0.95\hsize}{!}{\includegraphics{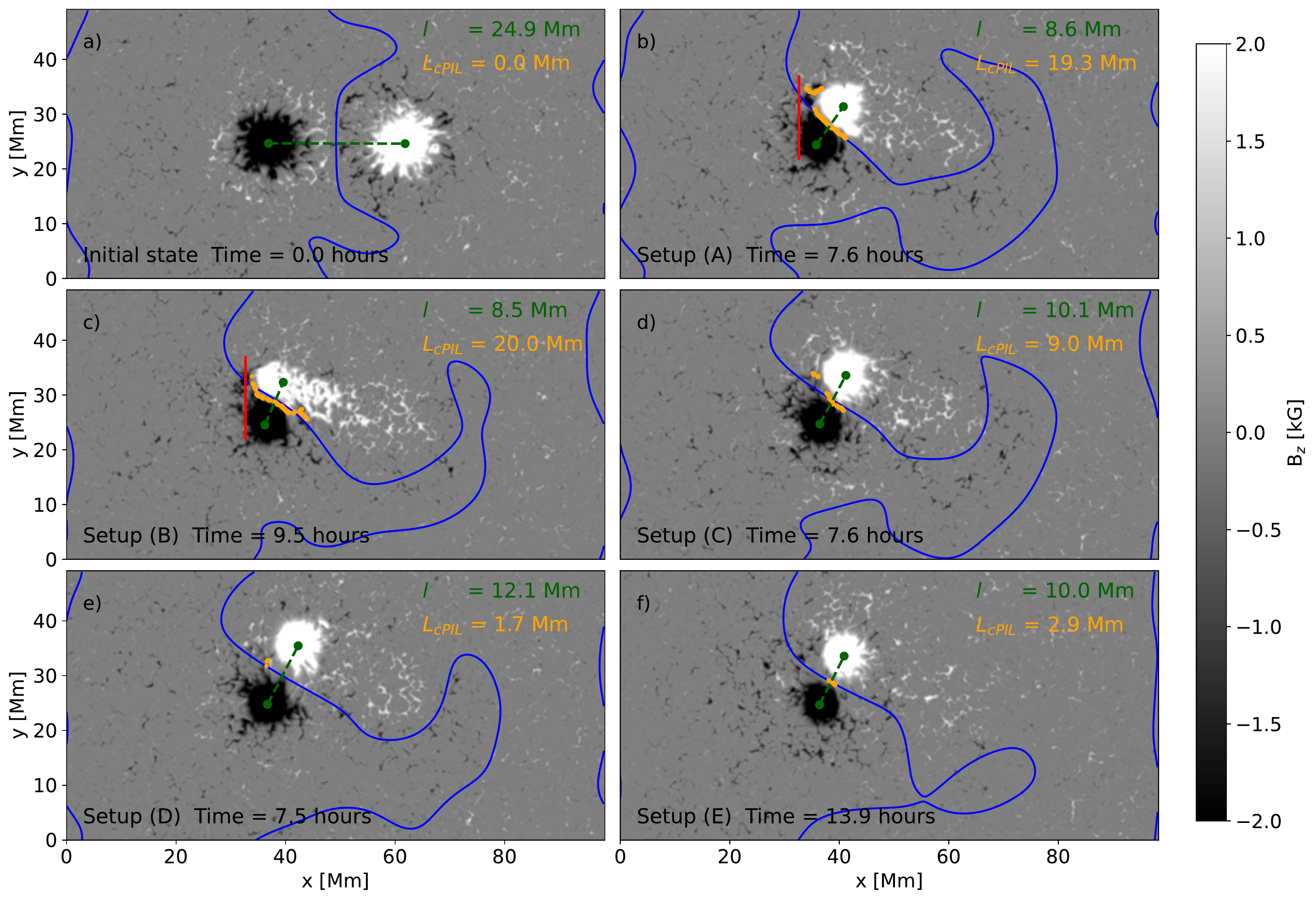}}
    \caption{Photospheric magnetograms \new{($\tau=0.1$)}. a) Initial state (identical for all setups); b-f) configuration found in setups (A)-(E) \new{at the time of closest encounter}. Setups (A) and (B) differ in terms of coherence of the moving spot and \new{are analyzed in detail in this present paper}. The red vertical line indicates the position of vertical cross-sections presented in Figures \ref{fig:section_A} and \ref{fig:section_B}. \new{The blue contour highlights the PIL based on a smoothed magnetogram, the orange contours the cPILs. The spot centroids and collision distance are indicated by green dots and a green dashed line, respectively.}}
    \label{fig:magnetogram}
\end{figure}

\begin{figure}
    \centering
    \resizebox{0.95\hsize}{!}{\includegraphics{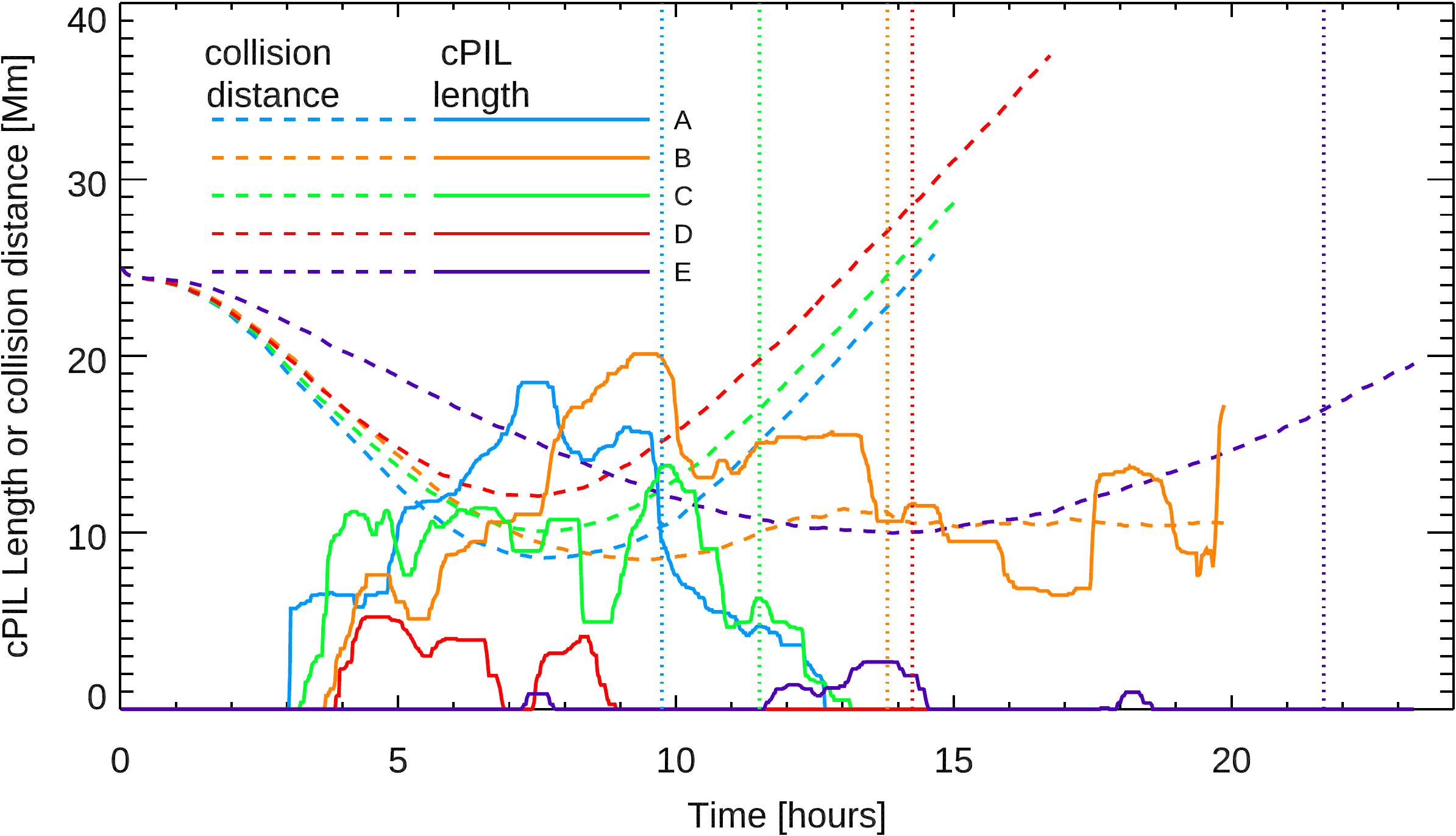}}
    \caption{\new{Time evolution of cPIL length and collision distance for setups (A)-(E). Vertical dotted lines indicate the peak time of the flare (or the peak time of the strongest flare in case there was more than one flare) that was produced in each setup. See text for discussion.}
    }
    \label{fig:cPIL_distance}
\end{figure}

\begin{figure}
    \centering
    \resizebox{0.95\hsize}{!}{\includegraphics{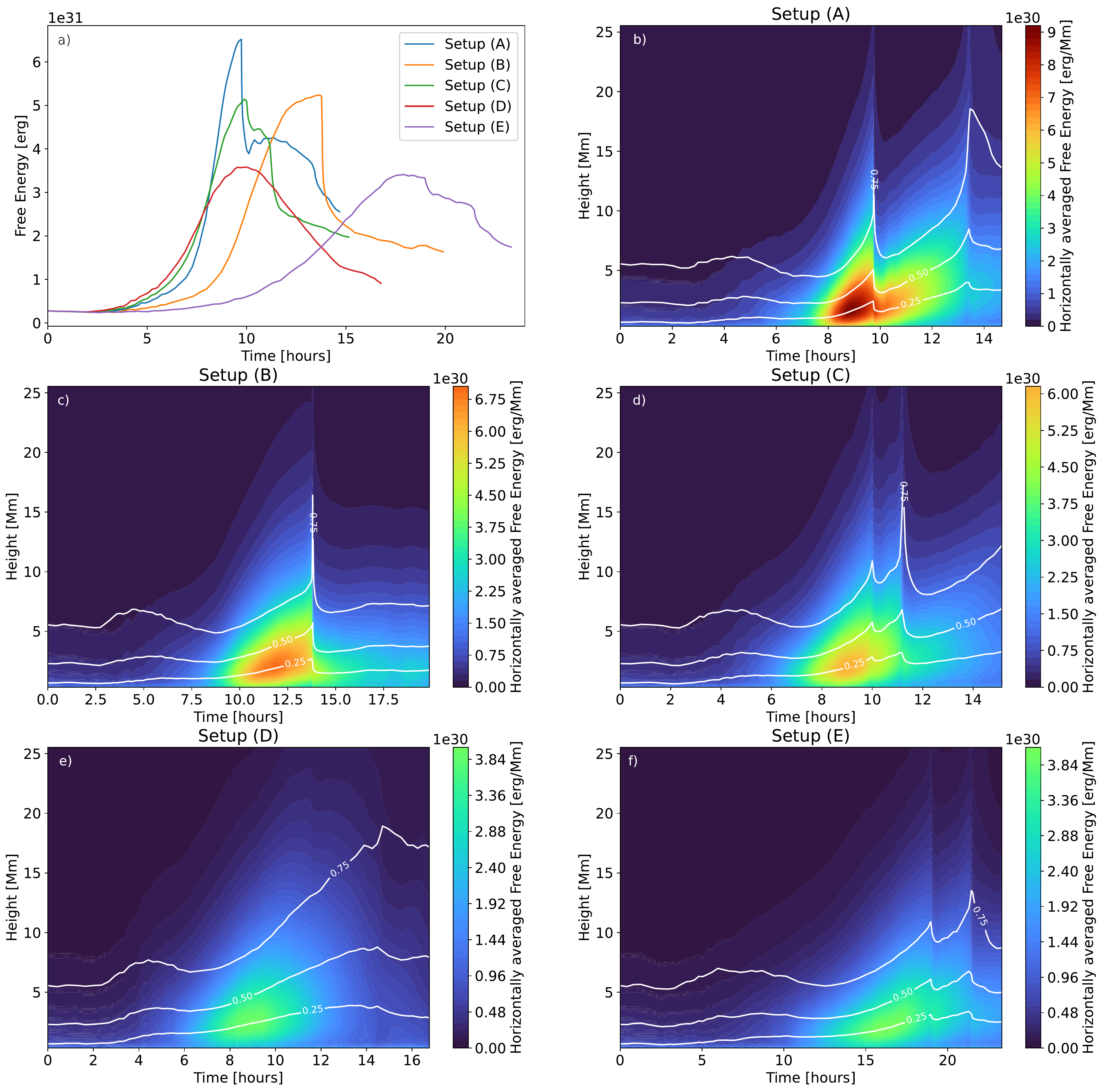}}
    \caption{Buildup of magnetic free energy in setups (A)-(E). Panel a) shows the time evolution of magnetic free energy integrated over the coronal volume of the simulation domain. Panels b-f) show the time evolution of the horizontally averaged magnetic free energy for the individual cases. Here white contour lines indicate the height levels beneath which $25/50/75\%$ of the magnetic free energy is stored.}
    \label{fig:free-energy}
\end{figure}

\begin{figure}
    \centering
    \resizebox{0.95\hsize}{!}{\includegraphics{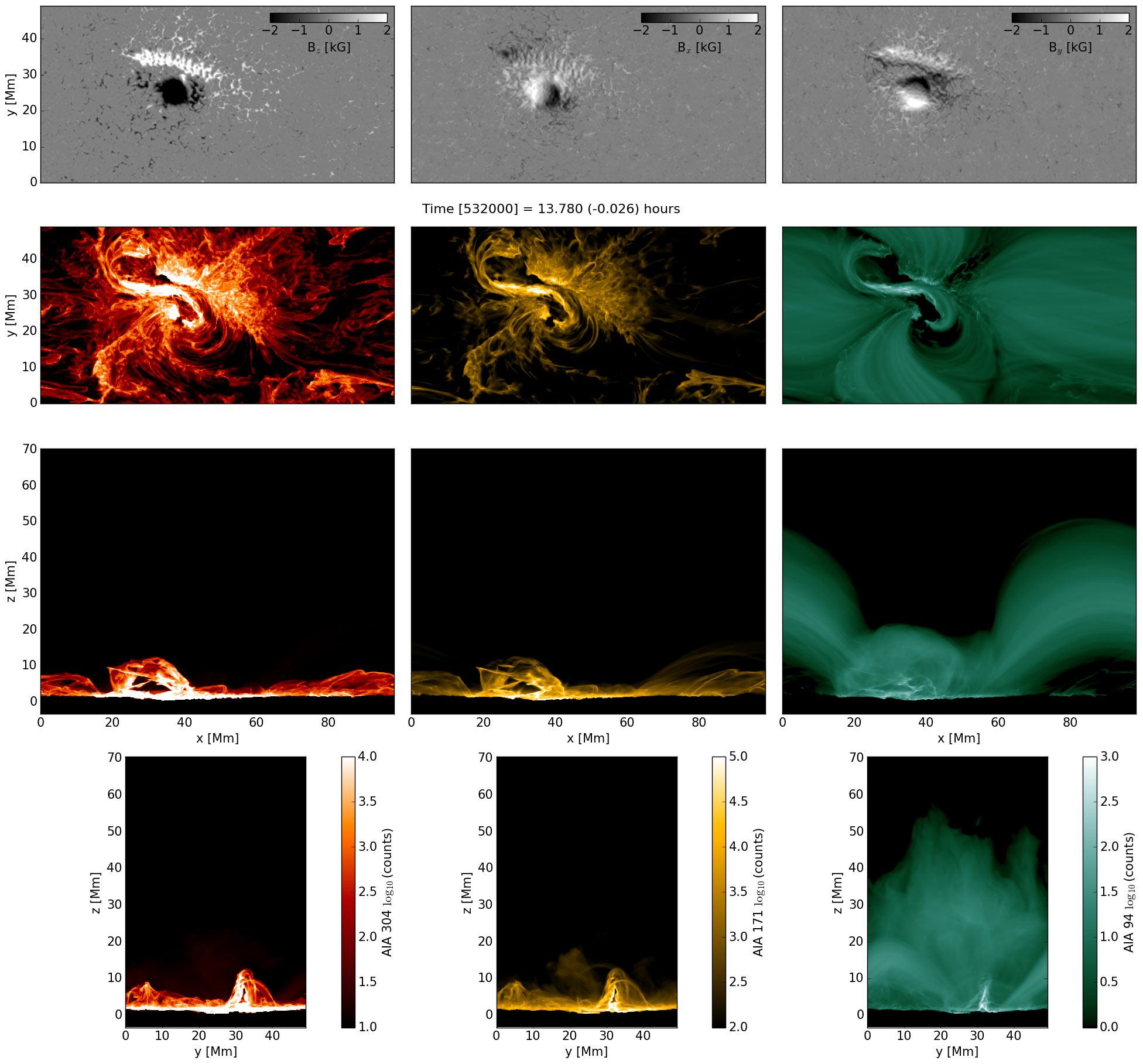}}
    \caption{Photospheric and coronal appearance for setup (B). Top: Magnetic field in photosphere extracted on the $\tau=0.1$ level. Bottom panels: \new{synthetic EUV emission in the He II 304\,\AA\ (left; sensitive to solar plasma temperature $\approx$ 50,000 K), Fe IX 171\,\AA\ (middle; $\approx$ 600,000 K) and Fe XVIII 94\,\AA\ (right; $\approx$ 6,300,000 K)} channels of SDO/AIA computed for the three grid-aligned view directions as indicated by the axis labels. The snapshot corresponds to a time \new{about 90 seconds} before the flare peak as defined by the GOES X-ray flux \new{(see Figure \ref{fig:goes_flux})}. The side views show in AIA 304\,\AA\ and 171\,\AA\ a prominence that will erupt during the flare. An animation of this figure is provided with the online material. \new{For better cross reference with other figures we indicate in this figure and animation the simulation time step, the time since the start of the simulation and the time relative to the flare peak as defined through GOES X-ray flux (see Figure \ref{fig:goes_flux}). The animation covers the full length of the simulation starting with the formation of an EUV sigmoid structure (before time step 470000), the formation and rise of a pre-eruption flux rope (time-steps 470000 - 530000), the impulsive phase of the flare to it's peak (time-steps 532000-5420000), the formation of post-flare arcade loops (time-steps 550000-600000), and the cooldown phase ending with coronal rain formation (time steps 750000-800000).}}
    \label{fig:HMI_AIA}
\end{figure}

\begin{figure}
    \centering
    \resizebox{0.95\hsize}{!}{\includegraphics{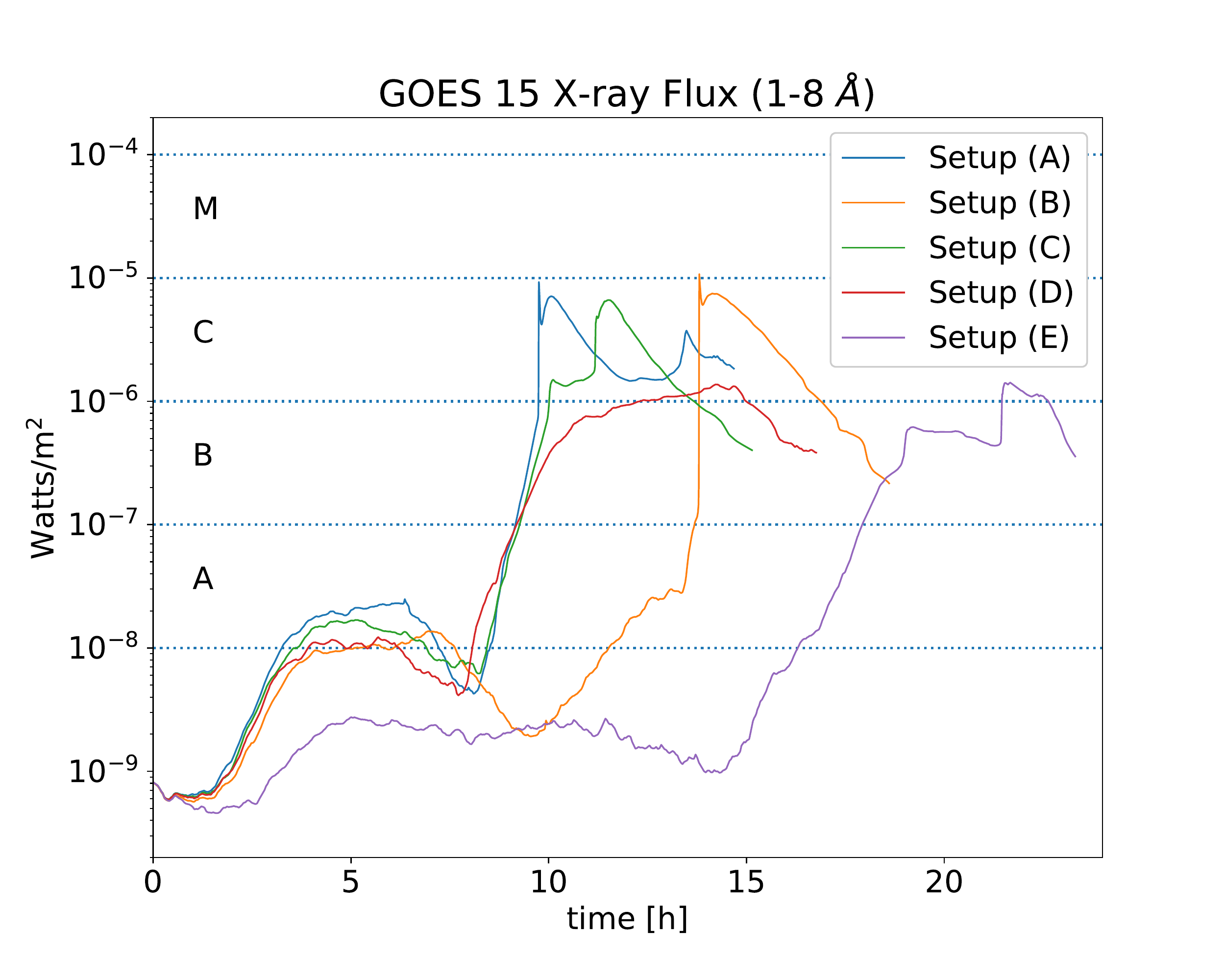}}
    \caption{Synthetic GOES 15 X-ray flux computed for setups (A)-(E). The two strongest flares found in setups (A) and (B) reach a GOES flux peak level of M1.0.}
    \label{fig:goes_flux}
\end{figure}

\begin{figure}
    \centering
    \resizebox{0.95\hsize}{!}{\includegraphics{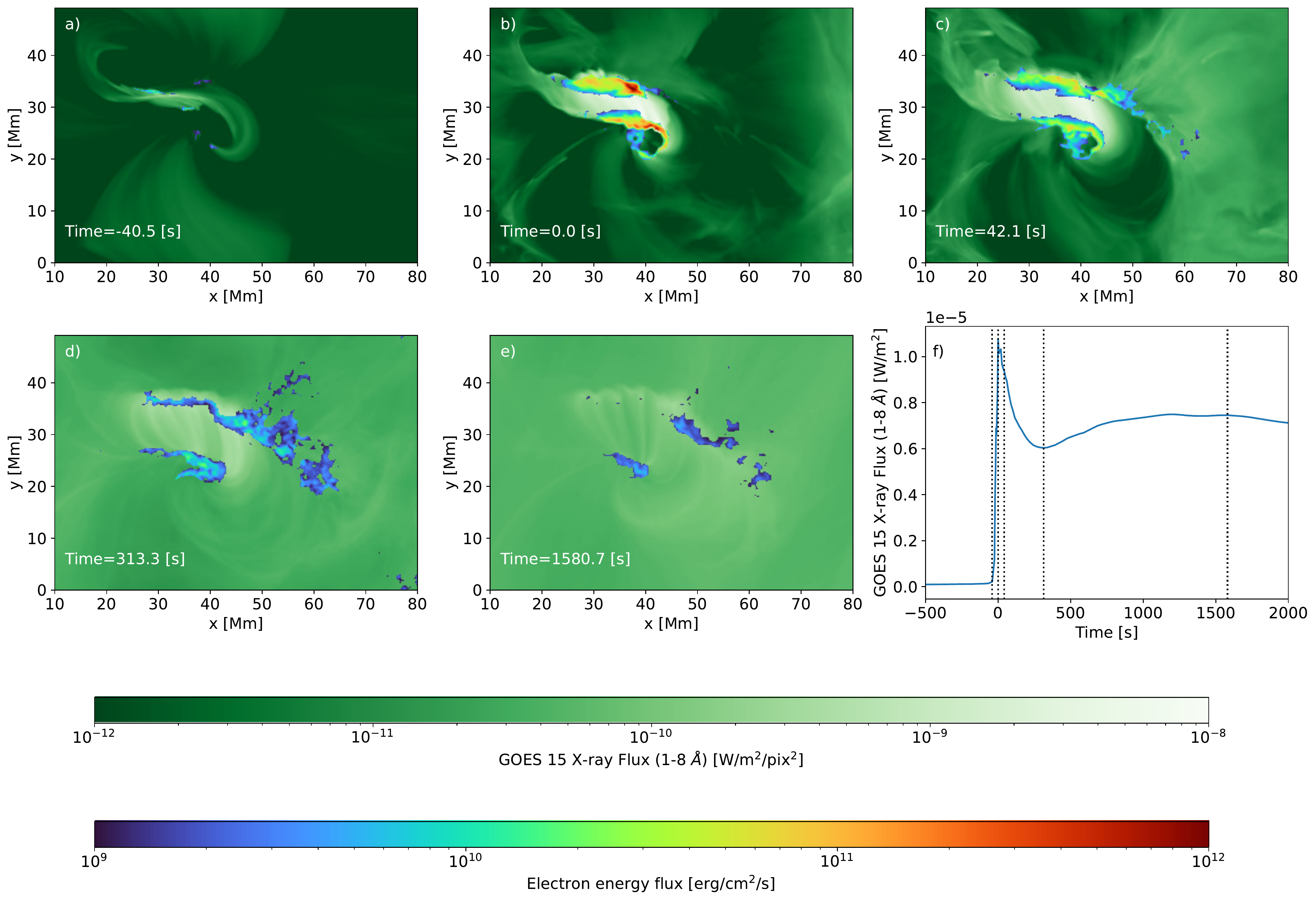}}
    \caption{Evolution for flare ribbons for setup (B). In panel a)-e) we present synthetic GOES 15 soft X-ray images \new{at 1-8\AA\ bandpass} with an overlay of the electron energy flux at a height of 1\,Mm above the average $\tau=1$ level in the simulation. Panel f) shows the integrated GOES 15 flux. Vertical dotted lines indicate the snapshots presented in panels a)-e). \new{The snapshots shown correspond to the timesteps 534000, 542000, 548000, 565000, and 600000 in the animation provided with Figure 4.}}
    \label{fig:ribbons}
\end{figure}

\begin{figure}
    \centering
    \resizebox{0.95\hsize}{!}{\includegraphics{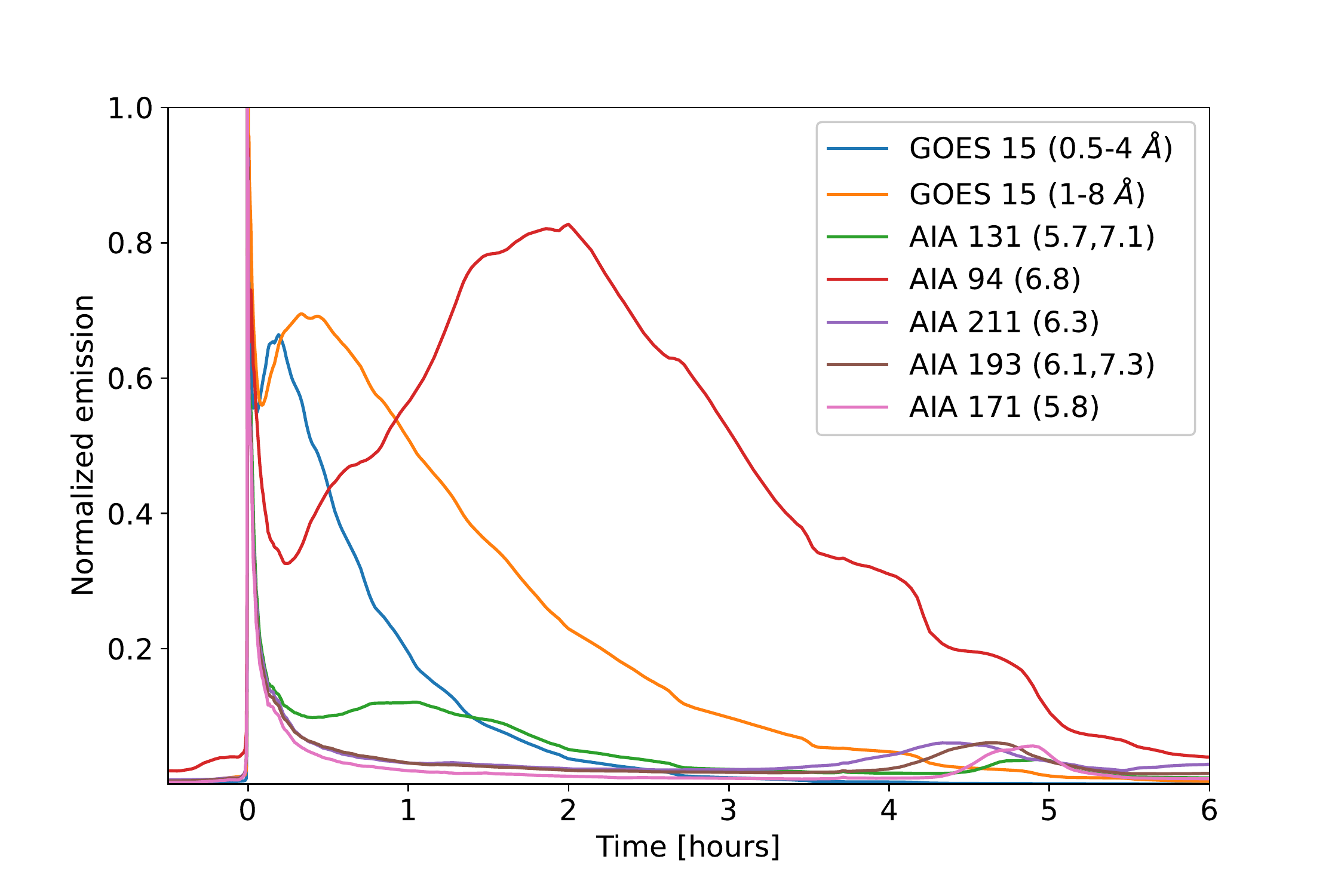}}
    \caption{Post-flare evolution of GOES 15 X-ray and AIA EUV flux for setup (B). The secondary peak is present in both X-ray and EUV flux and the time-delays are consistent with a typical cooling sequence. \new{We show for the AIA channels the logT of the dominant temperature contributions.}}
    \label{fig:Late_EUV}
\end{figure}

\subsection{General properties and energetics}

Figure \ref{fig:magnetogram} panel a) shows the initial photospheric magnetogram, $B_z$, \new{as the $\hat{z}$-component of the 3D MHD magnetic field solution, $\mathbf{B}_{MHD}=B_x\mathbf{\hat{x}}+B_y\mathbf{\hat{y}}+B_z\mathbf{\hat{z}}$ taken at the $\tau=0.1$ surface}\new{, which is the same for all setups,} and in panels b)-f) the magnetograms corresponding to the setups (A)-(E) at the time of minimum collision distance (i.e., during closest encounter; see exact times in Table~\ref{tab:tab1}). \new{In Figure \ref{fig:cPIL_distance}, we show the time evolution of the collision distance, $l$ (dashed lines), and the cPIL length, $L_{cPIL}$ (solid lines), for each setup. We measured the collision distance, $l$, by tracking each polarity's flux-weighted centroid with time by using a circular mask to isolate the dominant polarity and eliminate bias from other flux elements, as described in \citet{Chintzoglou:etal:2019}. $L_{cPIL}$ was measured after performing a few simple manipulations as described below: }

\new{(1) We convolved each frame of the $B_z$ image series with an assumed ``instrumental'' Gaussian Point Spread Function (PSF) of 1$''$ FWHM (assuming 1$'' \approx$ 720 km on the Sun, this means a Gaussian kernel of 3.75 pixels FWHM in the simulated $B_z$ image) and then resampled the $B_z$ maps of degraded resolution to simulate ``observed magnetograms'' from the Helioseismic and Magnetic Imager (HMI; \citealt{Schou:etal:2012}) instrument of the \emph{Solar Dynamics Observatory} (SDO; \citealt{Pesnell:etal:2012}) at a pixel scale of $\approx$ 0.36 Mm $pix^{-1}$ (the latter being the pixel scale of remapped SDO/HMI magnetograms to a Cylindrical Equal Area projection, as done in \citealt{Chintzoglou:etal:2019}). This particular manipulation allows us to apply the methodology of \citet{Chintzoglou:etal:2019} directly, as the latter was developed on magnetograms obtained by SDO/HMI. However, the strong-flux small-scale Moving Mangetic Features (MMFs; see a ring-like distribution of small-scale features of polarity opposite to that of the spots in Figure~\ref{fig:magnetogram}a) that move radially outwards to all radial directions from the simulated spots, caused many false-positive detections of cPILs (especially in the area external to the PIL, i.e., in the non-PIL-facing areas around the spots) which can bias the measurement of $L_{cPIL}$. Such challenge was not reported in \citet{Chintzoglou:etal:2019}, likely because in the HMI observations (of reduced spatial resolution as compared to our simulations) the HMI's instrumental PSF possibly smears out the appearance and magnetic strength of MMFs. In addition, the strength and size of MMFs might be more pronounced in these simulations as compared to real MMFs, also making the straightforward measurement of the $L_{cPIL}$ a challenging task. It is therefore necessary to mask the non-PIL-facing areas around the spots to properly focus in the area between the spots.}

\new{(2) To overcome the aforementioned challenge, we constructed time-dependent binary masks that were produced by the following morphological operations: (i) we produced bitmaps at a threshold of $|B_z|=2000\,G$ (to exclude most of the MMFs below that threshold) of a (boxcar) temporally-smoothed magnetogram series (used a time window of 10 frames to suppress any transient variability due to rapid MMF motions), and then (ii) performed image dilation of those bitmaps (for both the positive and negative polarity) using a square $30\times 30\ pix^2$ kernel, to finally (iii) extract the desired region between the dilated positive and negative polarity bitmaps as a result of their overlapping bitmap area. The combination of (1) and (2) (i)-(iii) enabled us to isolate the collision area at each time frame, by eliminating severe contamination from the MMFs around the spots. We then applied the cPIL detection methodology of \citet{Chintzoglou:etal:2019} in the extracted area between the moving polarities and measured $L_{cPIL}$ as a function of time for each setup. Lastly, we applied, in post-processing, a temporal median filter to the $L_{cPIL}$ measurements using a one-hour-long time window, to further suppress noise in the measurements introduced by any remaining MMFs (some occurring as the polarities approach/separate before/after the moment of minimum collision distance). The final results are shown in Figure~\ref{fig:cPIL_distance}.}

\new{We note a lag between the time of minimum collision distance, $t_{min}$ (the latter coinciding with maximum cPIL length for setups with smallest collision distances, i.e., for cases A and B) and flare peak time, $t_{peak}$ (vertical dotted lines in Figure~\ref{fig:cPIL_distance}). This seems to be in agreement with the observations, as the majority of activity produced over a cPIL appears to happen after the minimum collision distance is reached (see \citealt{Chintzoglou:etal:2019}, Figures 7d and 10d therein).} 

In this paper we primarily focus on setups (A) and (B) shown in panels b) and c) in Figure~\ref{fig:magnetogram}. These two setups have a similar collision speed and similar \new{angle of attack for the positive spot's motion in the x-y plane} and only differ in the coherence of the moving positive polarity spot. At the time of collision the spot in setup (B) is dispersed over a length of $\approx$ 20~Mm, leading to a \new{long, and persistent cPIL (in contrast, setup A's cPIL lasts less as the moving polarity maintains its cohesion and doesn't deform to produce a long and persistent cPIL)}. While both simulations, as discussed in detail below, do have a flare of comparable strength, there are distinct differences in the pre-eruption magnetic field configuration. In Figure~\ref{fig:magnetogram} panels b) and c) we highlight with a vertical red line the position of cross-sections we discuss further in Figures \ref{fig:section_A} and \ref{fig:section_B}.

\new{Figure \ref{fig:free-energy} a) shows the buildup and release of the magnetic free energy, $E_f$, defined as:}

\begin{equation}
    E_f(t)=\int \frac{\vec{B}_{\rm MHD}^2(x,y,z,t)-\vec{B}_{\rm pot}^2(x,y,z,t)}{8\pi} dV
\end{equation}

\noindent \new{where $\mathbf{B}_{\rm pot}$ the horizontally periodic potential field that is computed from the photospheric magnetogram of the MHD simulation, and $\mathbf{B}_{\rm MHD}$ the MHD magnetic field solution. The integral is taken over the volume, $V=\int_0^{L_x}\int_0^{L_y}\int_{h_{phot}}^{L_z} dxdydz$, starting with a grid layer in the photosphere corresponding to the average geometric height at $\tau=0.2$, $z=h_{phot}$, and ending at the top of the simulation domain.}

All cases \new{reach magnetic} free energies in the range of $3-6\times 10^{31}$~ergs. For the coherent spot setups (A), (C), and (D), the peak magnetic free energy is found shortly after the closest encounter of the spots around $t=9$ hours, for setup (E) with half the collision speed around $t=17$ hours. The significantly lower energy release in setup (E) in comparison to setup (C) is likely due to 2 reasons: (1) the slower collisional speed allows for more coronal energy loss due to numerical resistivity during the buildup phase; (2) both polarities show a continuous decay in the photosphere. The longer time to collision leads to spots with less flux and a larger effective separation during collision (which can be seen in Figure \ref{fig:magnetogram}). In the case of the dispersing spot (B) the magnetic free energy buildup is delayed by about 3~hours compared to setups (A), (C), and (D). Comparing setups (A) and (B) we find a comparable energy release of $2.5\times 10^{31}$ ergs during the flare, although setup (A) reaches, with about $6.5\times 10^{31}$ ergs, a higher value of magnetic free energy than setup (B) with about $5.2\times 10^{31}$~ergs, which makes these \emph{flares rather efficient in terms of releasing $40-50\%$ of the magnetic free energy \new{accumulated in the system}}. Setups (C) and (E) release energy in 2 smaller flares, setup (D) leads to a continuous energy release through multiple small-scale reconnection events. Figure \ref{fig:free-energy}, panels b)-e) \new{show for each setup the height distribution of (horizontally averaged) magnetic free energy as function of time, }

\begin{equation}
    <E_f(z ,t)>_{horizontal}=\int_0^{L_x}\int_0^{L_y} dx dy \frac{\vec{B}_{\rm MHD}^2(x,y,z,t)-\vec{B}_{\rm pot}^2(x,y,z,t)}{8\pi L_x L_y} ; \ \ \ h_{phot}\leq z\leq L_z
\end{equation}

\noindent \new{with overlaid height contours under which $25/50/75\%$ of the magnetic free energy is accumulated in each case. }In all cases $50\%$ of the magnetic free energy is stored below a height of $5$~Mm at the time when flares occur. 

Figure \ref{fig:HMI_AIA} shows a simulation snapshot from setup (B) at a time of \new{about 90 seconds} before the flare peak. We show the photospheric magnetic field and synthetic \new{EUV emission at 3 different channels (He II 304\,\AA, Fe IX 171\,\AA, and Fe XVIII 94\,\AA) of the Atmospheric Imaging Assembly (AIA; \citealt{Lemen:etal:2012}) instrument on board \emph{SDO}, from 3 different view points (one top-down view simulating a solar disk center view, and two side views simulating two alternative views at the solar limb)}. Specific to setup (B) is the formation of a filament channel that holds cool material and is visible in AIA 304\,\AA\ and 171\,\AA.

Figure \ref{fig:goes_flux} shows the synthetic X-ray flux for the \new{\emph{Geostationary Operational Environmental Satellite 15} (in short, GOES 15)} computed from the full simulation domain for the setups (A)-(E). \new{We chose to simulate X-ray fluxes from GOES 15 as the latter operated over a large part of the previous solar cycle and thus was used extensively to characterize the flare activity in ARs.} Setup (A) leads to a C9.0 flare followed by a weaker secondary C4.0 flare, setup (B) leads to a single M1.0 flare whereas setup (C) has a combination of C1.0 and a later C7.0 flare. Setups (D) and (E) reach C1.0 level, in the case of (D) without a distinct flare. The strong flares in setups (A) and (B) show a distinct double peak: a sharp peak at the end of impulsive phase and a more broad peak 20-30 minutes later. To our knowledge this has not been observed in real data, but, as we describe below, this behavior is similar to what is seen in late-phase EUV flares \citep{Woods:etal:2011:lateEUV}.

\begin{figure}
    \centering
    \resizebox{0.95\hsize}{!}{
    \includegraphics{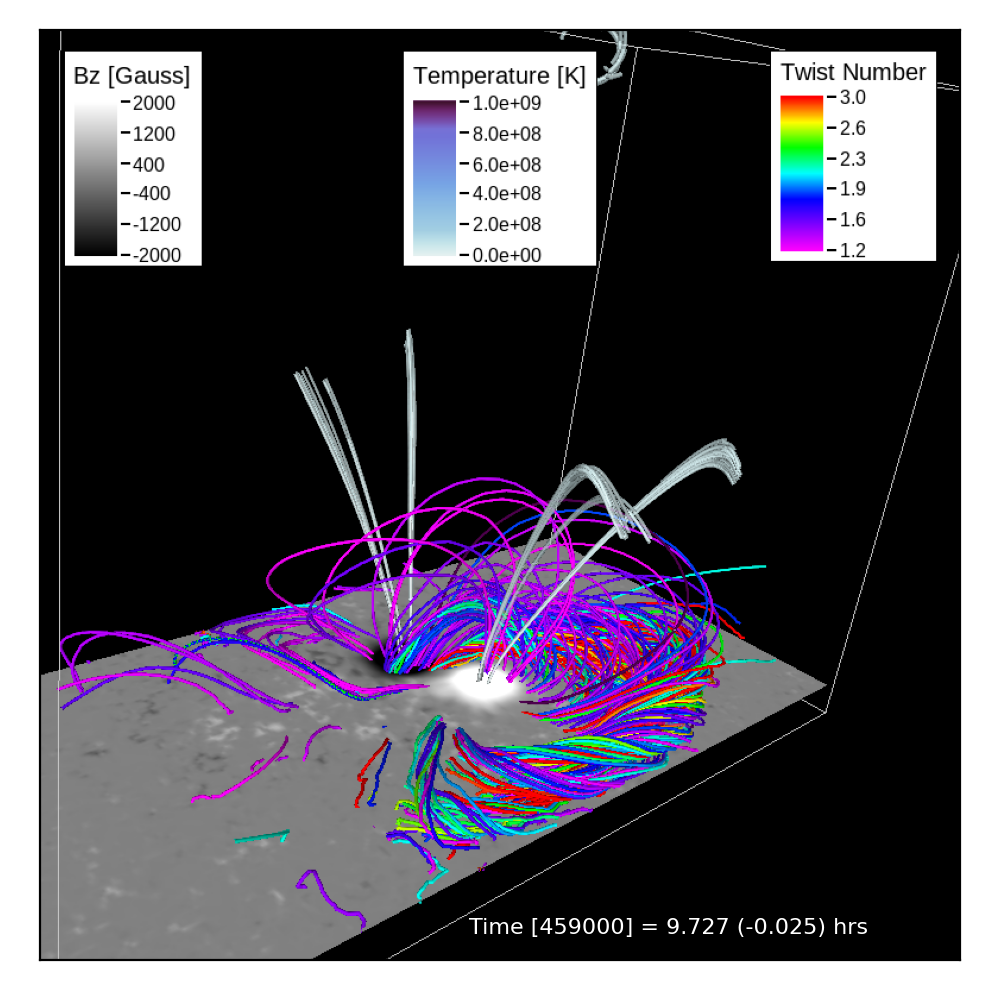}
    \includegraphics{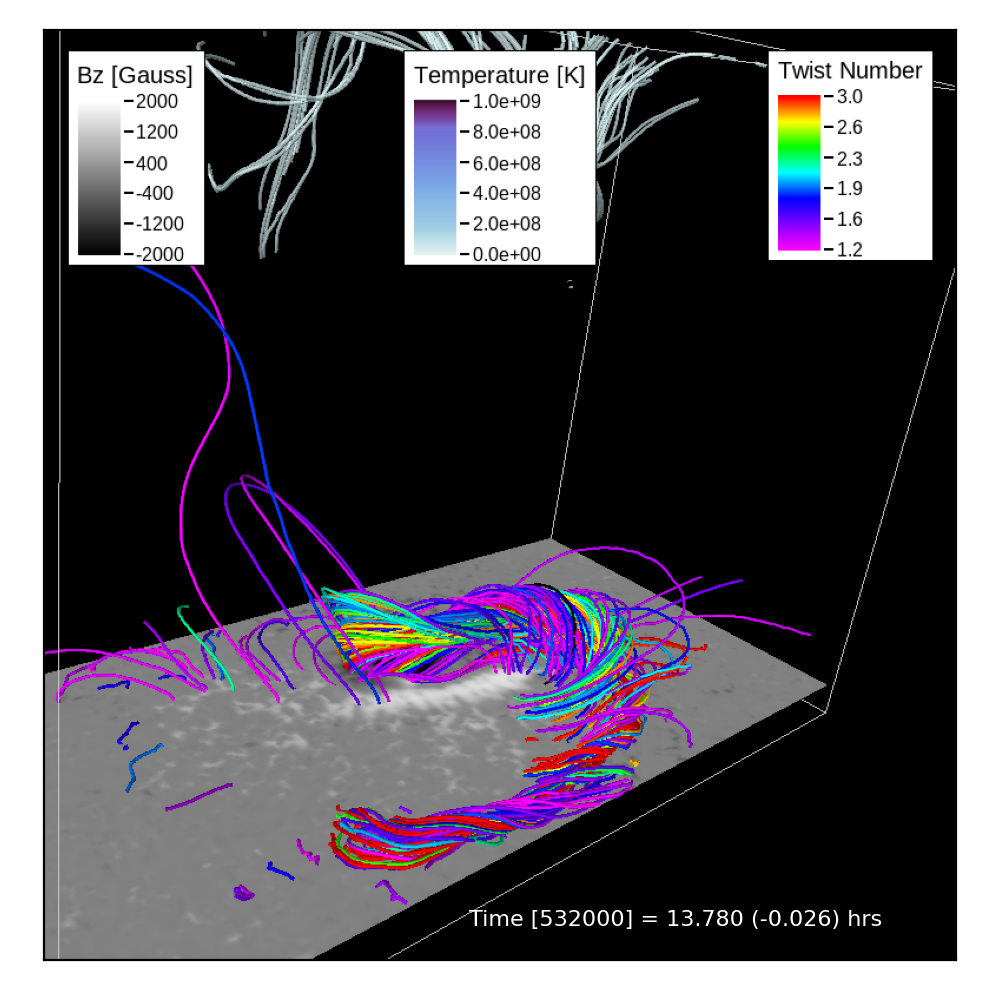}}
    \caption{Pre-eruptive magnetic field configuration \new{about 90 seconds} before flare peaks for setup (A) (left) and setup (B) (right). Magnetic field lines are color-coded according to their twist number, only field lines with values larger than 1.2 are shown. \new{In addition we show a second set of field lines which are selected based on their connectivity to the reconnection region, these field lines are color-coded by temperature.} The magnetogram shows the vertical component of the the magnetic field at the average position of the photosphere. An animation is provided in the online material. \new{The animation shows the evolution of the quantities displayed in this figure from the start of the simulation past the peak of the flares. A later snapshot of the movie is given in Figure \ref{fig:eruption-CME}.} Imagery produced by VAPOR (www.vapor.ucar.edu) \citep{2019Atmos..10..488L:vapor}.}
    \label{fig:pre-eruption}
\end{figure}

\begin{figure}
    \centering
    \resizebox{0.95\hsize}{!}{
    \includegraphics{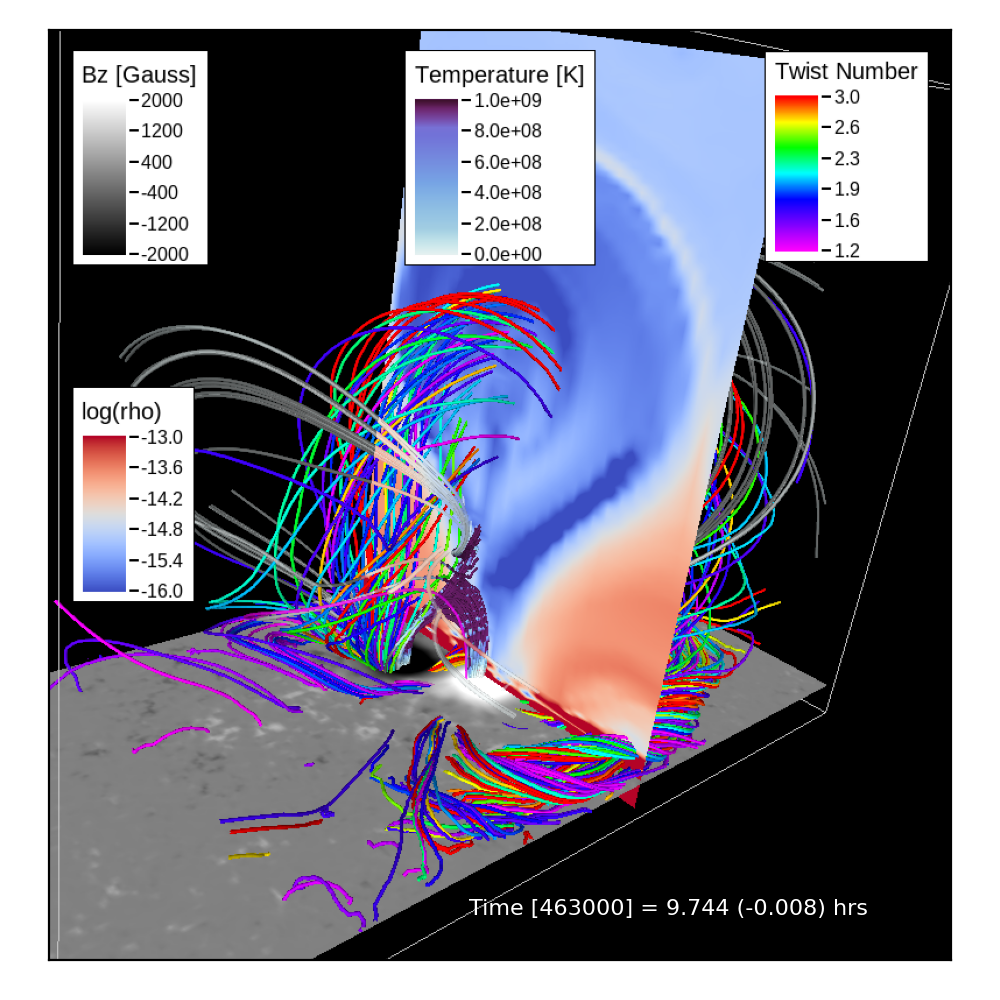}
    \includegraphics{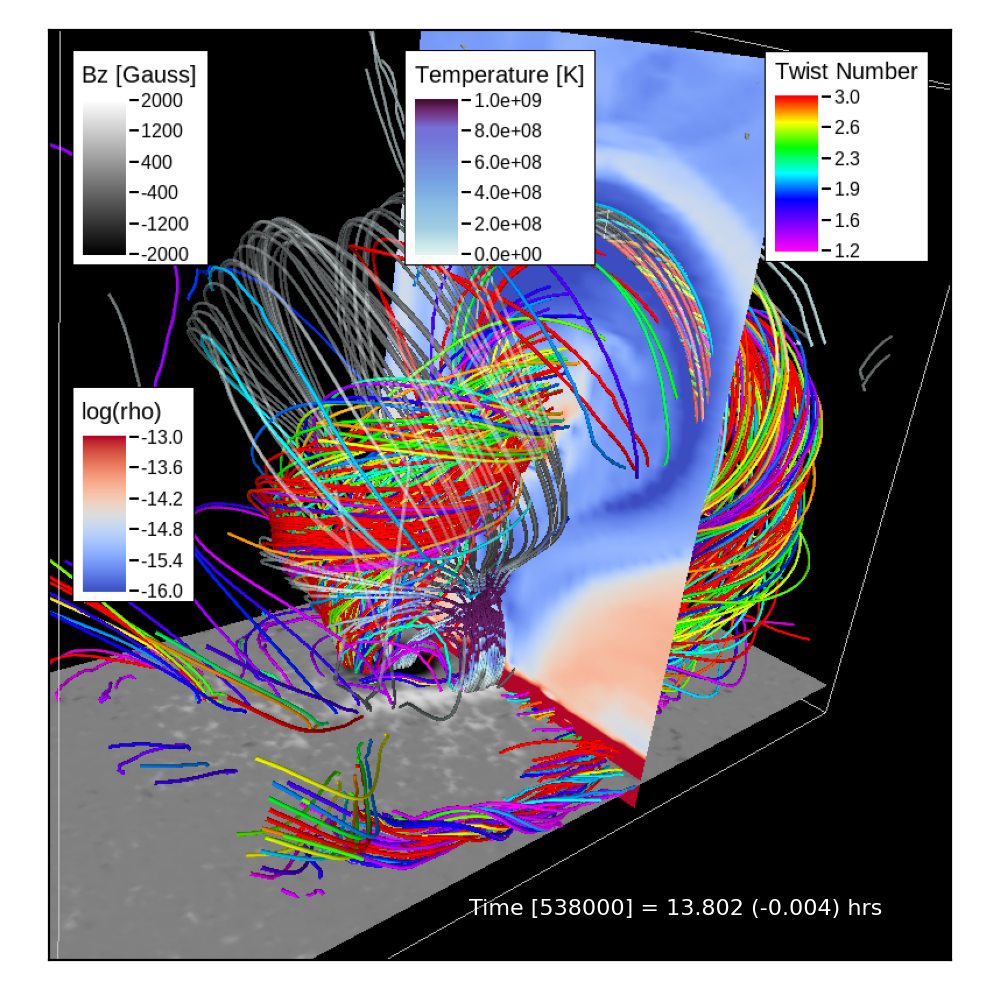}}
    \caption{Magnetic field configuration \new{28 seconds} before the flare peak for setup (A) (left) and \new{14 seconds before the flare peak for} setup (B) (right).  Field lines are color-coded by twist number as in Figure \ref{fig:pre-eruption}. In addition we show a second set of field lines which are selected based on their connectivity to the reconnection region, these field lines are color-coded by temperature. In both setups we find tether-cutting reconnection underneath the ejected flux-ropes. We show cross-sections through the erupting flux-ropes that display the mass density. Imagery produced by VAPOR (www.vapor.ucar.edu).}
    \label{fig:eruption-CME}
\end{figure}

\begin{figure}
    \centering
    \resizebox{0.95\hsize}{!}{\includegraphics{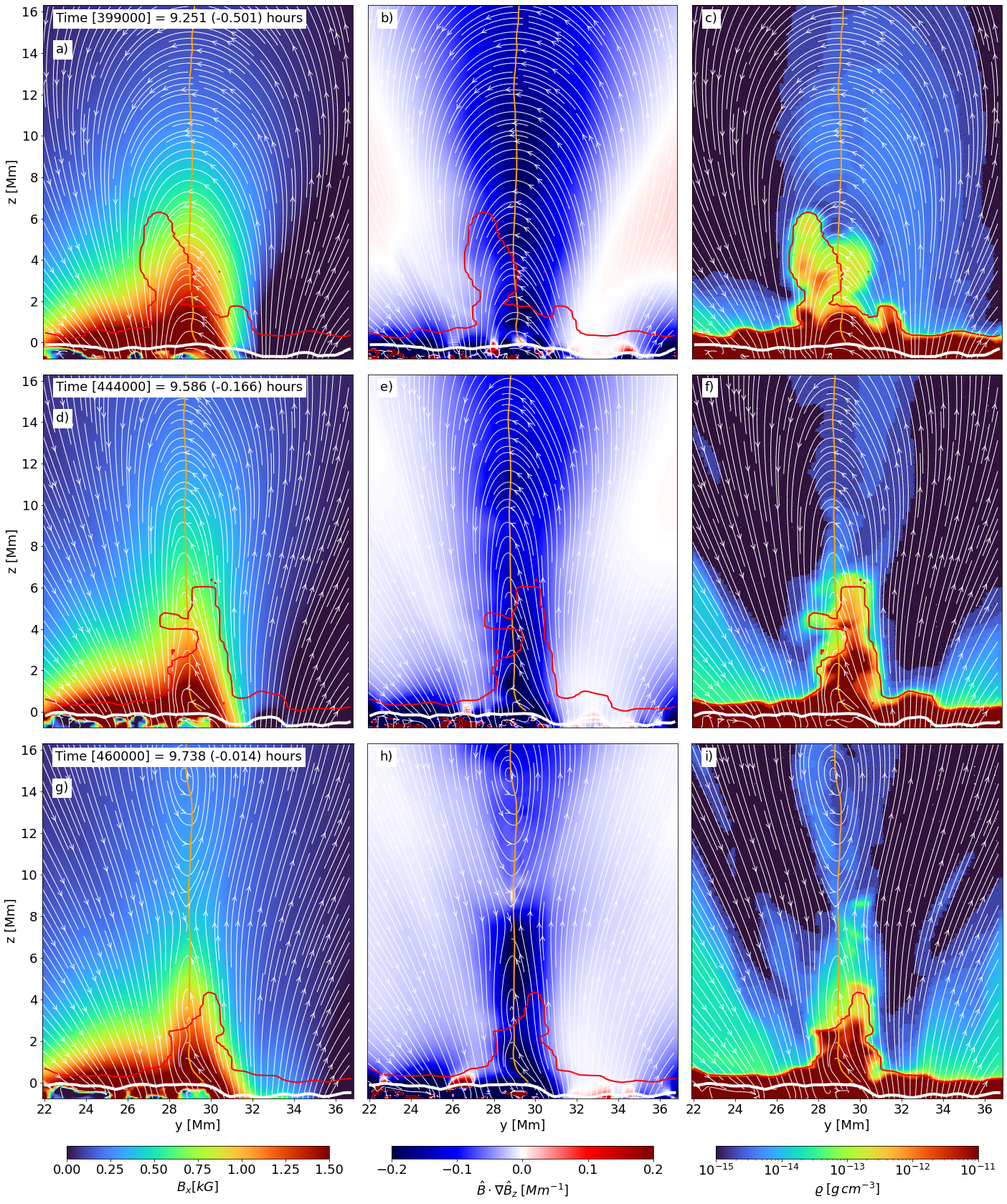}}
    \caption{Cross section from setup (A) along the cut indicated in Figure \ref{fig:magnetogram}b). Left panels show the \new{strength of $B_x$ (field component along PIL)}, middle panels the \new{vertical component of the} field line curvature and right panels the mass density. Top to bottom we show the time evolution as indicated. The  red line indicates the $T=50,000$~K contour, the orange line the $B_z=0$ contour. Field lines of the magnetic field within the y-z plane are shown in white. \new{The thick white contour of the corrugated $\tau=0.1$ level is plotted to indicate the photosphere (at $y>30$\,Mm the coherent and strong positive polarity causes slight Wilson depression of the $\tau$ surface).} A MFR carrying very little mass is formed mostly \new{during} the eruption.}
    \label{fig:section_A}
\end{figure}

\begin{figure}
    \centering
    \resizebox{0.95\hsize}{!}{\includegraphics{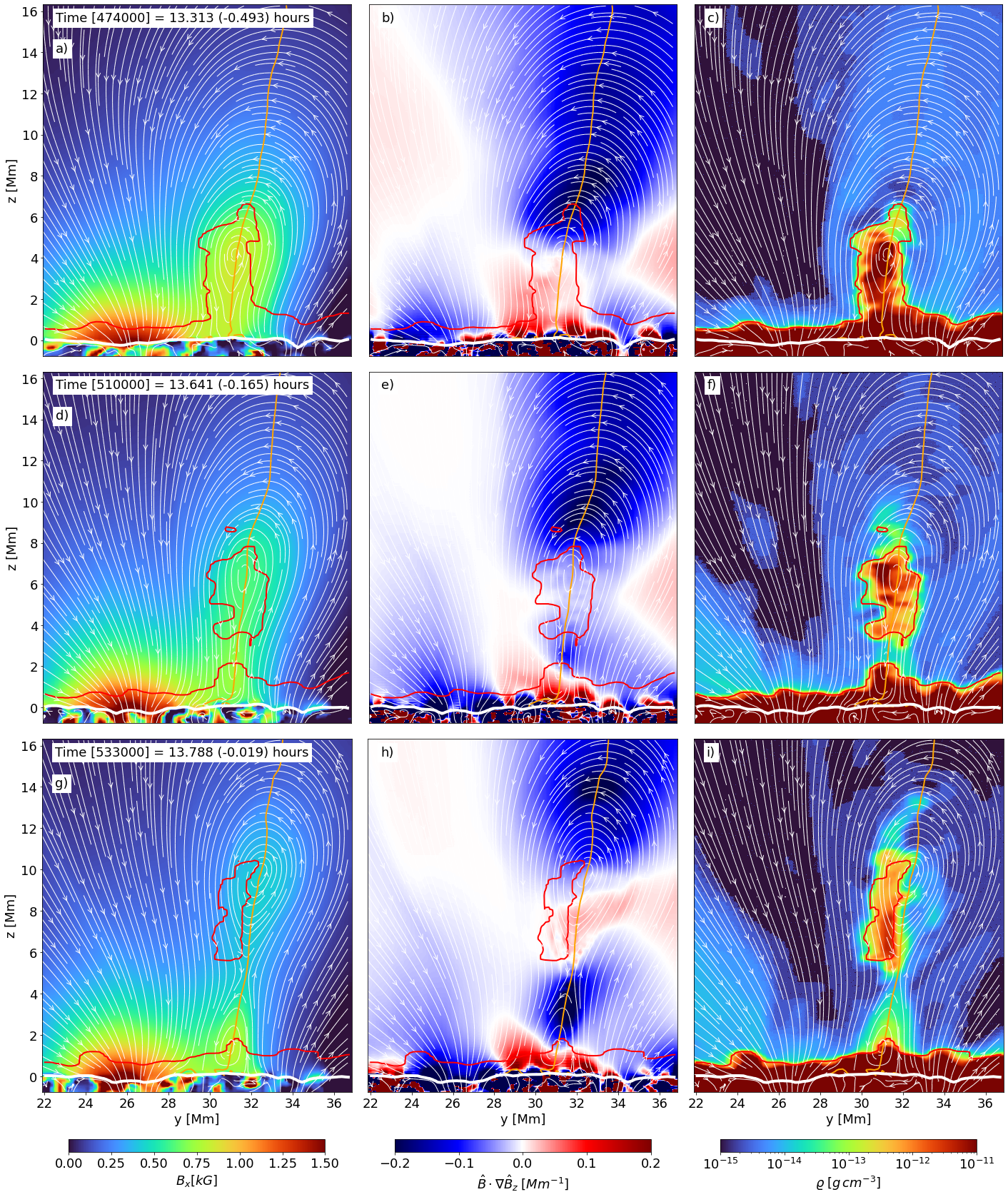}}
    \caption{Cross section from setup (B) along the cut indicated in Figure \ref{fig:magnetogram}c). All quantities shown are as in Figure \ref{fig:section_A}. A MFR is present already before the eruption and carries most of the filament mass away after the eruption.}
    \label{fig:section_B}
\end{figure}

\new{In Figures \ref{fig:HMI_AIA}, \ref{fig:ribbons}, \ref{fig:pre-eruption}, \ref{fig:eruption-CME}, \ref{fig:section_A} and \ref{fig:section_B} we provide the time relative to the GOES flare peak. We note that the onset of the eruption pre-dates the peak flare times for setups (A) and (B) by about 50 and 70 seconds, respectively.}

We further investigate the origin of the two GOES \new{1-8\,\AA} emission peaks in Figure \ref{fig:ribbons} where we show in panels a) - e) a map of the synthetic GOES flux together with the electron energy flux about $1$~Mm above the average $\tau=1$ level, which indicates the position of the flare ribbons. The electron energy flux is given by Spitzer conductivity that is limited during flare conditions by the free-streaming flux according to \citet{Fisher:etal:1985:flare}. The zero point in time corresponds to the peak on GOES X-ray flux as shown in panel e). $40.5$ seconds before the flare peak (panel a) the GOES X-ray flux imaging shows an X-ray sigmoid with enhancements of electron energy flux reaching $10^{10}$ ergs cm$^{-1}$ s$^{-1}$ in close proximity (onset of reconnection). During flare peak (panel b) the GOES flux highlights the hot post-flare loops, the electron energy flux reaches peak values of more than $10^{12}$ ergs cm$^{-1}$ s$^{-1}$ in small areas. $42.1$ seconds after the flare (panel c) the post flare loops and flare ribbons start fading, while diffuse X-ray emission outside the central region of the flare starts to increase and to dominate about $313.3$ seconds after the flare (panel d). The secondary peak is found about $25$ minutes after the flare and is dominated by diffuse emission from most of the simulation volume (panel e).

The high electron energy fluxes found in this simulation are a consequence of energy release in a rather small volume. As discussed above, about $2.5\times 10^{31}$ ergs are released, most of that in heights of less than $5$~Mm. We find reconnection outflows of up to $10,000$~km s$^{-1}$, leading to peak plasma energy densities of about $0.5$~MeV per particle (which would correspond to more than a billion K if interpreted as a thermal plasma). Our single fluid MHD approach cannot properly treat the exact partition of released energy between electrons and protons and expected deviations from Maxwellian distributions. However, a high energy density is in this setup unavoidable on the basis of energy conservation MHD is built upon. The secondary peak in the GOES 15 X-ray flux is consequence of this very hot plasma cooling down and passing through the GOES \new{1--8\,\AA} response about $25$ minutes after the flare. This becomes more evident from Figure \ref{fig:Late_EUV}, where we show in addition the shorter wavelength GOES channel \new{(0.5--4\,\AA)} as well as simulated AIA 131\,\AA, 94\,\AA, 211\,\AA, 193\,\AA\ and 171\,\AA. The secondary peak appears first in the $0.5-4$~{\AA} GOES channel followed by the $1-8$~{\AA} channel. About one hour after the flare peak we see a secondary peak in AIA 131\,\AA\ (this pass band has contributions around logT$\sim 5.7$ and $7.1$), followed by the secondary AIA 94\,\AA\ peak about 2 hours later. The pass bands of AIA 211\,\AA, 193\,\AA\ and 171\,\AA\ show a secondary peak about $4-5$~hours after the flare. Note that there is a tertiary peak of AIA 131\,\AA\ at around $5$~hours, which results from the logT$\sim 5.7$ contribution to this passband. This simulated flare is possibly an example for a late-phase EUV flare as described in \citet{Woods:etal:2011:lateEUV}. In this type of flares a secondary peak in EUV emission originating not from the vicinity of the flare site is found minutes to hours after the flare. 

\subsection{Evolution of magnetic field topology}
Figure \ref{fig:pre-eruption} shows the pre-eruption magnetic field structure \new{about 90 seconds before the eruption} for setup (A) and (B) in panels a) and b), respectively. \new{We highlight in Figure \ref{fig:pre-eruption} magnetic field lines that are associated with strong field-aligned currents. To that end we select and color code field lines by the quantity,} \new{$T_w$, known as the \emph{Twist number}, defined as} :
\begin{equation}
    T_w=\frac{1}{4\,\pi}\int\frac{(\nabla\times\vec{B})\cdot\vec{B}}{\vert\vec{B}\vert^2}{\rm d}s
\end{equation}

\noindent \new{where, $\mathbf{B}$, the magnetic field vector, and $ds$, the infinitesimal length of a magnetic field line.} \new{We note that $T_w$ does not provide an accurate measure for the number of windings  of field lines around the axis of a flux rope, but is generally non-linearly related to the winding number \citep[see, e.g.][]{Moffatt:Ricca:1992}, which can be computed after a proper identification of the flux-rope axis. We refer the reader also to the discussion in the Appendix C of \citet{Liu:etal:2016:twist}, about the relation of $T_w$ to the more general winding number $T_g$.}

We only show field lines that have twist numbers above 1.2. While both setups do host strongly twisted field lines prior to the eruption, only setup (B) has a MFR with dipped field lines that can support a pre-eruption filament. While the magnetic field structure in setup (A) is more arcade like (see also Fig. \ref{fig:section_A}), the presence of twisted field lines suggests a hybrid state between SMA and MFR. This difference becomes evident in Figure \ref{fig:eruption-CME}, \new{which shows the erupting MFR for setups (A) and (B) during the onset of eruption about 28 and 14 seconds before the flare peak, respectively.} The MFR in setup (B) is more strongly twisted and does carry more mass in its core. In addition to the twist number we highlight in Fig.~\ref{fig:eruption-CME} field lines that are  connected to the reconnection region. To this end we use random seeds with a bias towards high temperature and color-code the field lines by their temperature. In both setups we find tether-cutting reconnection in a similar location. As discussed earlier, the ``temperature'' exceeds in these setups a billion K due to the single fluid approach that cannot separate between thermal and non-thermal particles. We find energies per particle up to 0.5 MeV at the reconnection sites.

Figures \ref{fig:section_A} and \ref{fig:section_B} show cross sections for setups (A) and (B) at the positions indicated in Figure \ref{fig:magnetogram} by red lines in panels b) and c). We show \new{$B_x$ on the left}, the \new{vertical component of the field line curvature $\vec{\hat{B}}\cdot\vec{\nabla} \hat{B}_z$, with $\hat{\vec{B}}=\vec{B}/\vert\vec{B}\vert$,} in the middle and the mass density, $\varrho$, on the right. In order to delineate cool chromospheric and filament material from the hot corona we show the $T=50,000$~K contour in red. The contour of $B_z=0$~G is shown in orange, while white field lines indicate the field components within the y-z plane. For setup (A) the magnetic field strength is monotonically decreasing with height for the two snapshots prior to the eruption. The field line curvature is negative throughout, indicating the absence of dips that can hold filament mass. Nonetheless cool material is present above the cPIL and as a consequence the transition region is elevated to about 6 Mm height. A flux rope, indicated by closed field lines in the y-z plane and the presence of positive field line curvature is only present after the \new{eruption}. Since the region with positive field line curvature (dipped field lines) formed above the region with enhanced mass density (at a height of about $7-8$~Mm), the erupting MFR carries only very little mass. 

The situation is very different for setup (B), shown in Figure \ref{fig:section_B}. Here a MFR with dipped field lines (positive curvature) is already present more than half an hour before the flare. The reconnection forming this MFR happens deep in the atmosphere close to photospheric heights and therefore results in a larger amount of filament mass supported by dipped field lines. As a consequence of the close to force-free field configuration ($(\vec{B}\cdot\vec{\nabla})\vec{B}=\frac{1}{2}\vec{\nabla}\vec{B}^2$) the magnetic field strength has a local maximum on the MFR axis and consequently \new{the magnetic field component $B_x$ (field along the PIL, which is the dominant contribution to $\vert \vec{B} \vert$ above the PIL)} (left panels) shows little variation with height above the PIL, which is a prominent difference to setup (A) with a monotonically decreasing field strength. About 10 minutes prior to the flare the axis of the MFR starts to rise and most of the cool material present above the PIL is lifted since it is supported by the dipped field lines. After the flare the erupting MFR carries most of the filament mass.

\section{Future observations needed to constrain the pre-eruption magnetic field configuration}
\label{sec:future-obs}
The differences between SMA and MFR (and any transitional hybrid state) are evident at chromospheric heights since in both cases cool plasma is present to a height of about 6~Mm prior to the eruption. The most striking difference is visible in the variation of $\vert\vec{B}\vert$ with height, which is in the case of the SMA dropping from about 2~kG in the photosphere to less than 1~kG in a height of 5~Mm, while in the case of the MFR the field strength stays around 750~G over the same height range. While \new{field strength values are certainly setup-dependent}, the presence (or absence) of a monotonic vertical gradient \new{appears as a} critical distinction between SMA and MFR as it is strongly linked to field line curvature. This difference is measurable with spectro-polarimetric chromospheric observations that do have a dense height coverage from photosphere to transition region. Recently, \citet{Judge:etal:2021} identified the near-UV spectral region near the \ion{Mg}{2} lines as suitable for this. The \ion{Mg}{2} diagnostics in the upper chromosphere can be complemented by a multitude of \ion{Fe}{1} and \ion{Fe}{2} lines that provide the height coverage. We note that for the distinction between SMA and MFR it is sufficient to analyze the variation of $\vert\vec{B}\vert$ with optical depth, a reconstruction on a physical height scale is, while certainly desirable, not strictly necessary since the difference between SMA and MFR is very striking. 
We strongly encourage the development of an instrument that can diagnose chromospheric magnetic field in ARs with dense height and continuous time coverage to capture the full evolution of cPILs.

\section{Conclusions} \label{sec:concl}
Collisional shearing \citep{Chintzoglou:etal:2019} is a common mechanism in complex flare- and CME-productive ARs \citep[see, also][]{Liu:etal:2019,Liu:etal:2021,Wang:etal:2022}. We studied a simplified setup that only considers the interaction of 2 opposite polarities. While this setup neglects the typically more complex quadrupolar magnetic field configuration present in observed explosive ARs, it does capture processes that happen near collisional polarity inversion lines (cPILs). Furthermore, this setup removes complexities originating from \new{active} flux emergence, as we simply translate the magnetic polarities, while still enabling flux cancellation in the photosphere. ARs with collisional shearing do often show a deformation of colliding spots that supports the formation of a longer cPIL 
\citep[e.g.,][]{Chintzoglou:etal:2019}. While we were able to partially emulate that effect in our setup (B) by allowing the moving spot to disperse, this is different from most observed behavior. Our spot dispersion is essentially spreading out the spot by leaving flux behind largely stationary, while it would be more realistic if these fragments continue to move and through that enhance the shearing and thereby increase the energy input into the corona.
\new{In Figure~\ref{fig:cPIL_distance} we have measured the  collision distance, $l$, and the collisional PIL length, $L_{cPIL}$, with time, for each of our (A)-(E) simulation setups. In all setups, we note a time-lag between the moment of minimum collision distance (the latter coinciding with maximum cPIL length for smallest collision distances, at least for setups A and B) and peak of major flare (vertical dotted lines in Figure~\ref{fig:cPIL_distance}; also see Table~\ref{tab:tab1}). This appears to be consistent with  observations of ARs undergoing collisional shearing, where the majority of activity over a cPIL tends to occur after the minimum collision distance is reached (see \citealt{Chintzoglou:etal:2019}, Figures 7d and 10d therein).} 

\new{The maximum length of the cPIL in setups (A) and (B) is comparable, however, the cPIL length \emph{persistence} is different. Setup (B) exhibits a longer (up to $\approx$20\,Mm long) cPIL for a longer time interval as compared to setup (A) (see Figure~\ref{fig:cPIL_distance}). This might suggest that more flux is able to cancel across a long cPIL that lasts for a much longer time (setup B) as compared to a similarly long cPIL which lasts for a shorter time (setup A). This may explain why setup (B) produced a denser eruptive filament as compared to setup (A). In the observations of \citet{Chintzoglou:etal:2019}, the cPILs of AR11158 and AR12017 persisted for several days, producing extreme flaring and CME activity.}

We studied the dependence \new{of explosive solar activity} on collision distance, collision speed and spot coherence \new{in our MHD simulations} and found:
\begin{enumerate}
    \item The collision distance, $l$, is the most critical parameter influencing the strength of the resulting flare(s). Only for the 2 closest collision setups we find \new{impulsive} flares, while in the more distant collision setup we find still a significant (but $\approx$ 2$\times$ lower) energy release in the form of enhanced heating distributed over multiple hours \new{(gradual low C-class activity, or lower)}. These findings are consistent with the idea of collisional shearing; a very small collision distance was also associated with spot deformation (indicating severity of collision) and explosive flare activity \citep{Chintzoglou:etal:2019}.
    \item Reducing the collision speed reduces the flare strength.
    \item The coherence of the spots during the collision and therefore the length of the resulting cPIL, $L_{cPIL}$, influences critically the pre-flare magnetic field configuration, specifically whether the transition from SMA to MFR happens before or during the flare.
    \item We find a similar flare energy release for both SMA and MFR setups, the primary difference is in the amount of mass that is ejected in the resulting CME \new{(higher amount of mass with pre-existing MFR, also considering that the MFR had formed over a more persistent cPIL, i.e., a long cPIL which lasted for a longer time interval}).
    \item The resulting lower M-class flares reach high energy densities of up to $0.5$ ~MeV per particle and electron energy flux in excess of $10^{12}~\mbox{erg}~\mbox{cm}^{-2}~\mbox{s}^{-1}$. Since the resulting hot corona takes several hours to cool, they produce late-phase EUV emission.
    \item Overall we find the flares resulting from collisional Polarity Inversion Lines (cPILs) \emph{very efficient}, i.e., up to $40-50\%$ of the stored magnetic free energy is released in the setups with close encounter (i.e., setups A, B, C). \new{For comparison, the flare simulation described in \citet{Cheung:etal:2019:HGCR} (parasitic flux emergence) released only about $10\%$ of the free magnetic energy.}
    \item Spectro-polarimetric observations in the chromosphere that provide continuous height coverage from photosphere to upper chromosphere could provide direct measurements of magnetic field that allow to distinguish between SMAs and MFRs prior to eruption.
\end{enumerate}

Our findings support that collisional shearing is a process that is effective in creating energetic eruptions. While our setup did not produce X-Class flares, it does produce significantly more energetic eruptions (up to a M1.0 flare) than the setup in \citet{Cheung:etal:2019:HGCR} that simulated a flare arising from parasitic flux emergence. While our simulation domain and AR size (magnetic flux of $3\times 10^{21}$~Mx) are similar to \citet{Cheung:etal:2019:HGCR}, our most energetic flares in setups (A) and (B) release 5 times more energy ($2.5\times 10^{31}$ ergs vs $5\times 10^{30}$ ergs), resulting in coronal energy densities and electron energy fluxes to the flare ribbons about an order of magnitude larger. If we extrapolate our setup to the case of NOAA AR11158 (the colliding polarities had magnetic fluxes of $6-8\times 10^{21}$~Mx during the collision process; \citealt{Chintzoglou:etal:2019}), we would expect that an energy release exceeding $10^{32}$~ergs \new{(X-class flare regime)} is feasible. We find electron fluxes in excess of $10^{12}$ ergs cm$^{-2}$ s$^{-1}$, which are towards the upper end of fluxes typically needed to explain flare ribbon properties in solar flares \citep{Kowalski:etal:2017,Kowalski:etal:2022}. The high fluxes and energy densities are a consequence of having a rather confined flaring volume in this setup, which is also the cause for late-phase EUV emission as described in \citet{Woods:etal:2011:lateEUV}. Owing to the high temperatures reached in the flaring corona, it takes the plasma a time span of several hours to cool back to typical coronal temperatures. Synthesizing several AIA channels we find EUV flux enhancements several hours after the flare corresponding to progressing cooling sequence. 

\begin{acknowledgements}
We thank the anonymous referee for constructive suggestions all of which helped to improve the final form of the manuscript. M.R. and G.C. are partially funded through NASA award 80NSSC19K0855 ``Investigating the Physical Processes Leading to Major Solar Activity''. G.C. also acknowledges partial support by NASA contract NNG04EA00C (SDO/AIA). Computing time was provided through NASA HECC resources under compute project s2169. Y.F. acknowledges partial support by the NASA LWS grant 80NSSC19K0070. L.K. is supported by a SNSF PRIMA grant.
This material is based upon work supported by the National Center for Atmospheric Research, which is a major facility sponsored by the National Science Foundation under Cooperative Agreement No. 1852977. We would like to acknowledge high-performance computing support from Cheyenne (doi:10.5065/D6RX99HX) provided by NCAR's Computational and Information Systems Laboratory, sponsored by the National Science Foundation. VAPOR is a product of the National Center for Atmospheric Research’s Computational and Information Systems Lab. Support for VAPOR is provided by the U.S. National Science Foundation (grants 03-25934 and 09-06379, ACI-14-40412), and by the Korea Institute of Science and Technology Information. M.R. is grateful to Stanislaw Jaroszynski and John Clyne who implemented custom changes that enable twist number calculation within VAPOR. CSIRO acknowledges the traditional owners of the land, sea, and waters, of the areas that we live and work on across Australia.
\end{acknowledgements}

\bibliography{papref_m}

\begin{thebibliography}{}
\expandafter\ifx\csname natexlab\endcsname\relax\def\natexlab#1{#1}\fi
\providecommand{\url}[1]{\href{#1}{#1}}
\providecommand{\dodoi}[1]{doi:~\href{http://doi.org/#1}{\nolinkurl{#1}}}
\providecommand{\doeprint}[1]{\href{http://ascl.net/#1}{\nolinkurl{http://ascl.net/#1}}}
\providecommand{\doarXiv}[1]{\href{https://arxiv.org/abs/#1}{\nolinkurl{https://arxiv.org/abs/#1}}}

\bibitem[{{Amari} {et~al.}(2000){Amari}, {Luciani}, {Mikic}, \&
  {Linker}}]{Amari:etal:2000}
{Amari}, T., {Luciani}, J.~F., {Mikic}, Z., \& {Linker}, J. 2000, \apjl, 529,
  L49, \dodoi{10.1086/312444}

\bibitem[{{Antiochos} {et~al.}(1999){Antiochos}, {DeVore}, \&
  {Klimchuk}}]{Antiochos:etal:1999}
{Antiochos}, S.~K., {DeVore}, C.~R., \& {Klimchuk}, J.~A. 1999, \apj, 510, 485,
  \dodoi{10.1086/306563}

\bibitem[{{Barnes} {et~al.}(2007){Barnes}, {Leka}, {Schumer}, \&
  {Della-Rose}}]{Barnes:etal:2007}
{Barnes}, G., {Leka}, K.~D., {Schumer}, E.~A., \& {Della-Rose}, D.~J. 2007,
  Space Weather, 5, S09002, \dodoi{10.1029/2007SW000317}

\bibitem[{{Cheung} {et~al.}(2019){Cheung}, {Rempel}, {Chintzoglou}, {Chen},
  {Testa}, {Mart{\'\i}nez-Sykora}, {Sainz Dalda}, {DeRosa}, {Malanushenko},
  {Hansteen}, {De Pontieu}, {Carlsson}, {Gudiksen}, \&
  {McIntosh}}]{Cheung:etal:2019:HGCR}
{Cheung}, M.~C.~M., {Rempel}, M., {Chintzoglou}, G., {et~al.} 2019, Nature
  Astronomy, 3, 160, \dodoi{10.1038/s41550-018-0629-3}

\bibitem[{{Cheung} {et~al.}(2022){Cheung}, {Mart{\'\i}nez-Sykora}, {Testa}, {De
  Pontieu}, {Chintzoglou}, {Rempel}, {Polito}, {Kerr}, {Reeves}, {Fletcher},
  {Jin}, {N{\'o}brega-Siverio}, {Danilovic}, {Antolin}, {Allred}, {Hansteen},
  {Ugarte-Urra}, {DeLuca}, {Longcope}, {Takasao}, {DeRosa}, {Boerner},
  {Jaeggli}, {Nitta}, {Daw}, {Carlsson}, {Golub}, \& {The}}]{Cheung:etal:2022}
{Cheung}, M. C.~M., {Mart{\'\i}nez-Sykora}, J., {Testa}, P., {et~al.} 2022,
  \apj, 926, 53, \dodoi{10.3847/1538-4357/ac4223}

\bibitem[{{Chintzoglou} \& {Zhang}(2013)}]{Chintzoglou:Zhang:2013}
{Chintzoglou}, G., \& {Zhang}, J. 2013, \apjl, 764, L3,
  \dodoi{10.1088/2041-8205/764/1/L3}

\bibitem[{{Chintzoglou} {et~al.}(2019){Chintzoglou}, {Zhang}, {Cheung}, \&
  {Kazachenko}}]{Chintzoglou:etal:2019}
{Chintzoglou}, G., {Zhang}, J., {Cheung}, M. C.~M., \& {Kazachenko}, M. 2019,
  \apj, 871, 67, \dodoi{10.3847/1538-4357/aaef30}

\bibitem[{{Cui} {et~al.}(2007){Cui}, {Li}, {Wang}, \& {He}}]{Yanmei:etal:2007}
{Cui}, Y., {Li}, R., {Wang}, H., \& {He}, H. 2007, \solphys, 242, 1,
  \dodoi{10.1007/s11207-007-0369-5}

\bibitem[{{Cui} {et~al.}(2006){Cui}, {Li}, {Zhang}, {He}, \&
  {Wang}}]{Yanmei:etal:2006}
{Cui}, Y., {Li}, R., {Zhang}, L., {He}, Y., \& {Wang}, H. 2006, \solphys, 237,
  45, \dodoi{10.1007/s11207-006-0077-6}

\bibitem[{{Fan}(2001)}]{Fan:2001}
{Fan}, Y. 2001, \apj, 546, 509.
\newblock
  \url{http://adsabs.harvard.edu/cgi-bin/nph-bib_query?bibcode=2001ApJ...546..509F&db_key=AST}

\bibitem[{{Fang} \& {Fan}(2015)}]{Fang:Fan:2015}
{Fang}, F., \& {Fan}, Y. 2015, \apj, 806, 79,
  \dodoi{10.1088/0004-637X/806/1/79}

\bibitem[{{Fisher} {et~al.}(1985){Fisher}, {Canfield}, \&
  {McClymont}}]{Fisher:etal:1985:flare}
{Fisher}, G.~H., {Canfield}, R.~C., \& {McClymont}, A.~N. 1985, \apj, 289, 414,
  \dodoi{10.1086/162901}

\bibitem[{{Gallagher} {et~al.}(2002){Gallagher}, {Moon}, \&
  {Wang}}]{Gallagher:etal:2002}
{Gallagher}, P.~T., {Moon}, Y.~J., \& {Wang}, H. 2002, \solphys, 209, 171,
  \dodoi{10.1023/A:1020950221179}

\bibitem[{{Georgoulis} \& {Rust}(2007)}]{Georgoulis:Rust:2007}
{Georgoulis}, M.~K., \& {Rust}, D.~M. 2007, \apjl, 661, L109,
  \dodoi{10.1086/518718}

\bibitem[{{Judge} {et~al.}(2021){Judge}, {Rempel}, {Ezzeddine}, {Kleint},
  {Egeland}, {Berdyugina}, {Berger}, {Bryans}, {Burkepile}, {Centeno}, {de
  Toma}, {Dikpati}, {Fan}, {Gilbert}, \& {Lacatus}}]{Judge:etal:2021}
{Judge}, P., {Rempel}, M., {Ezzeddine}, R., {et~al.} 2021, \apj, 917, 27,
  \dodoi{10.3847/1538-4357/ac081f}

\bibitem[{{Kliem} \& {T{\"o}r{\"o}k}(2006)}]{Kliem:Torok:2006}
{Kliem}, B., \& {T{\"o}r{\"o}k}, T. 2006, \prl, 96, 255002,
  \dodoi{10.1103/PhysRevLett.96.255002}

\bibitem[{{Kowalski} {et~al.}(2022){Kowalski}, {Allred}, {Carlsson}, {Kerr},
  {Tremblay}, {Namekata}, {Kuridze}, \& {Uitenbroek}}]{Kowalski:etal:2022}
{Kowalski}, A.~F., {Allred}, J.~C., {Carlsson}, M., {et~al.} 2022, \apj, 928,
  190, \dodoi{10.3847/1538-4357/ac5174}

\bibitem[{{Kowalski} {et~al.}(2017){Kowalski}, {Allred}, {Daw}, {Cauzzi}, \&
  {Carlsson}}]{Kowalski:etal:2017}
{Kowalski}, A.~F., {Allred}, J.~C., {Daw}, A., {Cauzzi}, G., \& {Carlsson}, M.
  2017, \apj, 836, 12, \dodoi{10.3847/1538-4357/836/1/12}

\bibitem[{{K{\"u}nzel}(1960)}]{Kuenzel:1960}
{K{\"u}nzel}, H. 1960, Astronomische Nachrichten, 285, 271,
  \dodoi{10.1002/asna.19592850516}

\bibitem[{{Lemen} {et~al.}(2012){Lemen}, {Title}, {Akin}, {Boerner}, {Chou},
  {Drake}, {Duncan}, {Edwards}, {Friedlaender}, {Heyman}, {Hurlburt}, {Katz},
  {Kushner}, {Levay}, {Lindgren}, {Mathur}, {McFeaters}, {Mitchell}, {Rehse},
  {Schrijver}, {Springer}, {Stern}, {Tarbell}, {Wuelser}, {Wolfson}, {Yanari},
  {Bookbinder}, {Cheimets}, {Caldwell}, {Deluca}, {Gates}, {Golub}, {Park},
  {Podgorski}, {Bush}, {Scherrer}, {Gummin}, {Smith}, {Auker}, {Jerram},
  {Pool}, {Soufli}, {Windt}, {Beardsley}, {Clapp}, {Lang}, \&
  {Waltham}}]{Lemen:etal:2012}
{Lemen}, J.~R., {Title}, A.~M., {Akin}, D.~J., {et~al.} 2012, \solphys, 275,
  17, \dodoi{10.1007/s11207-011-9776-8}

\bibitem[{{Li} {et~al.}(2019){Li}, {Jaroszynski}, {Pearse}, {Orf}, \&
  {Clyne}}]{2019Atmos..10..488L:vapor}
{Li}, S., {Jaroszynski}, S., {Pearse}, S., {Orf}, L., \& {Clyne}, J. 2019,
  Atmosphere, 10, 488, \dodoi{10.3390/atmos10090488}

\bibitem[{{Linton} {et~al.}(1998){Linton}, {Dahlburg}, {Fisher}, \&
  {Longcope}}]{Linton:etal:1998}
{Linton}, M.~G., {Dahlburg}, R.~B., {Fisher}, G.~H., \& {Longcope}, D.~W. 1998,
  \apj, 507, 404, \dodoi{10.1086/306299}

\bibitem[{{Linton} {et~al.}(1999){Linton}, {Fisher}, {Dahlburg}, \&
  {Fan}}]{Linton:etal:1999}
{Linton}, M.~G., {Fisher}, G.~H., {Dahlburg}, R.~B., \& {Fan}, Y. 1999, \apj,
  522, 1190, \dodoi{10.1086/307678}

\bibitem[{{Liu} {et~al.}(2019){Liu}, {Cheng}, {Wang}, \&
  {Zhou}}]{Liu:etal:2019}
{Liu}, L., {Cheng}, X., {Wang}, Y., \& {Zhou}, Z. 2019, \apj, 884, 45,
  \dodoi{10.3847/1538-4357/ab3c6c}

\bibitem[{{Liu} {et~al.}(2021){Liu}, {Wang}, {Zhou}, \& {Cui}}]{Liu:etal:2021}
{Liu}, L., {Wang}, Y., {Zhou}, Z., \& {Cui}, J. 2021, \apj, 909, 142,
  \dodoi{10.3847/1538-4357/abde37}

\bibitem[{{Liu} {et~al.}(2016){Liu}, {Kliem}, {Titov}, {Chen}, {Wang}, {Wang},
  {Liu}, {Xu}, \& {Wiegelmann}}]{Liu:etal:2016:twist}
{Liu}, R., {Kliem}, B., {Titov}, V.~S., {et~al.} 2016, \apj, 818, 148,
  \dodoi{10.3847/0004-637X/818/2/148}

\bibitem[{{Lynch} {et~al.}(2008){Lynch}, {Antiochos}, {DeVore}, {Luhmann}, \&
  {Zurbuchen}}]{Lynch:etal:2008}
{Lynch}, B.~J., {Antiochos}, S.~K., {DeVore}, C.~R., {Luhmann}, J.~G., \&
  {Zurbuchen}, T.~H. 2008, \apj, 683, 1192, \dodoi{10.1086/589738}

\bibitem[{{Moffatt} \& {Ricca}(1992)}]{Moffatt:Ricca:1992}
{Moffatt}, H.~K., \& {Ricca}, R.~L. 1992, Proceedings of the Royal Society of
  London Series A, 439, 411, \dodoi{10.1098/rspa.1992.0159}

\bibitem[{{Patsourakos} {et~al.}(2020){Patsourakos}, {Vourlidas},
  {T{\"o}r{\"o}k}, {Kliem}, {Antiochos}, {Archontis}, {Aulanier}, {Cheng},
  {Chintzoglou}, {Georgoulis}, {Green}, {Leake}, {Moore}, {Nindos}, {Syntelis},
  {Yardley}, {Yurchyshyn}, \& {Zhang}}]{Patsourakos:etal:2020}
{Patsourakos}, S., {Vourlidas}, A., {T{\"o}r{\"o}k}, T., {et~al.} 2020, \ssr,
  216, 131, \dodoi{10.1007/s11214-020-00757-9}

\bibitem[{{Pesnell} {et~al.}(2012){Pesnell}, {Thompson}, \&
  {Chamberlin}}]{Pesnell:etal:2012}
{Pesnell}, W.~D., {Thompson}, B.~J., \& {Chamberlin}, P.~C. 2012, \solphys,
  275, 3, \dodoi{10.1007/s11207-011-9841-3}

\bibitem[{{Rempel}(2012)}]{Rempel:2012:penumbra}
{Rempel}, M. 2012, \apj, 750, 62, \dodoi{10.1088/0004-637X/750/1/62}

\bibitem[{{Rempel}(2014)}]{Rempel:2014:SSD}
---. 2014, \apj, 789, 132, \dodoi{10.1088/0004-637X/789/2/132}

\bibitem[{{Rempel}(2017)}]{Rempel:2017:corona}
---. 2017, \apj, 834, 10, \dodoi{10.3847/1538-4357/834/1/10}

\bibitem[{{Schou} {et~al.}(2012){Schou}, {Scherrer}, {Bush}, {Wachter},
  {Couvidat}, {Rabello-Soares}, {Bogart}, {Hoeksema}, {Liu}, {Duvall}, {Akin},
  {Allard}, {Miles}, {Rairden}, {Shine}, {Tarbell}, {Title}, {Wolfson},
  {Elmore}, {Norton}, \& {Tomczyk}}]{Schou:etal:2012}
{Schou}, J., {Scherrer}, P.~H., {Bush}, R.~I., {et~al.} 2012, \solphys, 275,
  229, \dodoi{10.1007/s11207-011-9842-2}

\bibitem[{{Schrijver}(2007)}]{Schrijver:2007}
{Schrijver}, C.~J. 2007, \apjl, 655, L117, \dodoi{10.1086/511857}

\bibitem[{{Takasao} {et~al.}(2015){Takasao}, {Fan}, {Cheung}, \&
  {Shibata}}]{Takasao:etal:2015}
{Takasao}, S., {Fan}, Y., {Cheung}, M. C.~M., \& {Shibata}, K. 2015, \apj, 813,
  112, \dodoi{10.1088/0004-637X/813/2/112}

\bibitem[{{Toriumi} {et~al.}(2014){Toriumi}, {Iida}, {Kusano}, {Bamba}, \&
  {Imada}}]{Toriumi:etal:2014}
{Toriumi}, S., {Iida}, Y., {Kusano}, K., {Bamba}, Y., \& {Imada}, S. 2014,
  \solphys, 289, 3351, \dodoi{10.1007/s11207-014-0502-1}

\bibitem[{{Toriumi} \& {Takasao}(2017)}]{Toriumi:Takasao:2017}
{Toriumi}, S., \& {Takasao}, S. 2017, \apj, 850, 39,
  \dodoi{10.3847/1538-4357/aa95c2}

\bibitem[{{van Ballegooijen} \& {Martens}(1989)}]{Ballegooijen:Martens:1989}
{van Ballegooijen}, A.~A., \& {Martens}, P.~C.~H. 1989, \apj, 343, 971,
  \dodoi{10.1086/167766}

\bibitem[{{Wang} {et~al.}(2022){Wang}, {Liu}, {Yang}, \& {Hu}}]{Wang:etal:2022}
{Wang}, R., {Liu}, Y.~D., {Yang}, S., \& {Hu}, H. 2022, \apj, 925, 202,
  \dodoi{10.3847/1538-4357/ac3f35}

\bibitem[{{Woods} {et~al.}(2011){Woods}, {Hock}, {Eparvier}, {Jones},
  {Chamberlin}, {Klimchuk}, {Didkovsky}, {Judge}, {Mariska}, {Warren},
  {Schrijver}, {Webb}, {Bailey}, \& {Tobiska}}]{Woods:etal:2011:lateEUV}
{Woods}, T.~N., {Hock}, R., {Eparvier}, F., {et~al.} 2011, \apj, 739, 59,
  \dodoi{10.1088/0004-637X/739/2/59}

\bibitem[{{Zirin} \& {Liggett}(1987)}]{Zirin:Liggett:1987}
{Zirin}, H., \& {Liggett}, M.~A. 1987, \solphys, 113, 267,
  \dodoi{10.1007/BF00147707}

\end{thebibliography}
\bibliographystyle{aasjournal}

\end{document}